\documentclass[aps,prx,twocolumn,amsmath,amssymb,showpacs,superscriptaddress,notitlepage]{revtex4-1}
\usepackage{amsmath,amssymb,amsfonts,bm}
\usepackage{graphicx}
\usepackage{dcolumn}
\usepackage{mathrsfs}
\usepackage{bbold}
\usepackage{dsfont}
\usepackage{dcolumn}
\usepackage{epstopdf}
\usepackage[colorlinks=true,linkcolor=blue,citecolor=blue, urlcolor=blue,bookmarks=false]{hyperref}
\usepackage{changes}
\usepackage{textgreek}

\begin{document}

\title{Probing $\bm{k}$-Space Alternating Spin Polarization via the Anomalous Hall Effect}

\author{Rui Chen}
\affiliation{Department of Physics, Hubei University, Wuhan 430062, China}

\author{Zi-Ming Wang}
\affiliation{Department of Physics and Chongqing Key Laboratory for Strongly Coupled Physics, Chongqing University, Chongqing 400044, China}

\author{Hai-Peng Sun}
\affiliation{College of Engineering Physics, Shenzhen Technology University, 518118 Shenzhen, China}
\affiliation{Institute for Theoretical Physics and Astrophysics, University of W\"urzburg, 97074 W\"urzburg, Germany}

\author{Bin Zhou}
\affiliation{Department of Physics, Hubei University, Wuhan 430062, China}

\author{Dong-Hui Xu}\email[]{donghuixu@cqu.edu.cn}
\affiliation{Department of Physics and Chongqing Key Laboratory for Strongly Coupled Physics, Chongqing University, Chongqing 400044, China}
\affiliation{Center of Quantum Materials and Devices, Chongqing University, Chongqing 400044, China}

\begin{abstract}
Altermagnets represent a recently discovered class of collinear magnets, characterized by antiparallel neighboring magnetic moments and alternating-sign spin polarization in momentum-space~($\bm{k}$-space). However, experimental methods for probing the $\bm{k}$-space spin polarization in altermagnets remain limited. In this work, we propose an approach to address this challenge by interfacing an altermagnet with the surface of a topological insulator. The massless Dirac fermions on the topological insulator surface acquire a mass due to the time-reversal symmetry breaking. The local $\bm{k}$-space magnetic moment at the Dirac point directly determines both the sign and magnitude of this Dirac mass, resulting in an anomalous Hall effect. By measuring the Hall conductance, we can extract the local $\bm{k}$-space magnetic moment. Moreover, we can map the global magnetic moment distribution by tuning the Dirac point position using an in-plane magnetic field, thereby revealing the $\bm{k}$-space spin density of the altermagnet. This work establishes the Dirac fermion on the topological insulator surface as a sensitive probe for unveiling spin characters of altermagnets and those of other unconventional antiferromagnets.
\end{abstract}
\maketitle

{\color{blue}\emph{Introduction}.}---Altermagnets, characterized by their distinctive non-relativistic spin group symmetry, represent an emerging class of magnetic materials in which adjacent collinear magnetic moments alternate~\cite{Smejkal22PRX,Bai2024AFM,
vsmejkal2020crystal,Naka19NC,Ahn19PRB,
hayami2019momentum,Yuan2020Giant,mazin2021prediction,
ma2021multifunctional,
ifmmode2022Emerging,krempasky2024altermagnetic,sdongFeSb2}. Unlike conventional collinear antiferromagnets, altermagnets exhibit the 
nonrelativistic momentum-space~($\bm{k}$-space) spin polarization and splitting due to the time-reversal symmetry breaking, resulting in unique electronic transport phenomena~\cite{Nakaprb,shao2021spin,FengZX22NE,ifmmode2022Giant,Fernandestoplogical,zhang2024prl}. These distinctive properties, combined with their potential applications in spintronics, have made altermagnets a subject of intensive research. However, a definitive reproducible experimental detection of the altermagnetic state remains elusive~\cite{Keler2024NpjS}. While RuO$_2$ was initially considered a promising altermagnet candidate ~\cite{Feng2022NatElec,Zhou2024PRL,Karube2022PRL,
Fedchenko2024SciAdv,Hiraishi2024PRL}, recent studies have cast doubt on this classification~\cite{liu2024arXiv,Smolyanyuk2023arXiv}.
Consequently, developing reliable methods for detecting altermagnetic states remains an important challenge.

\begin{figure}[h!]
\centering
\includegraphics[width=\columnwidth]{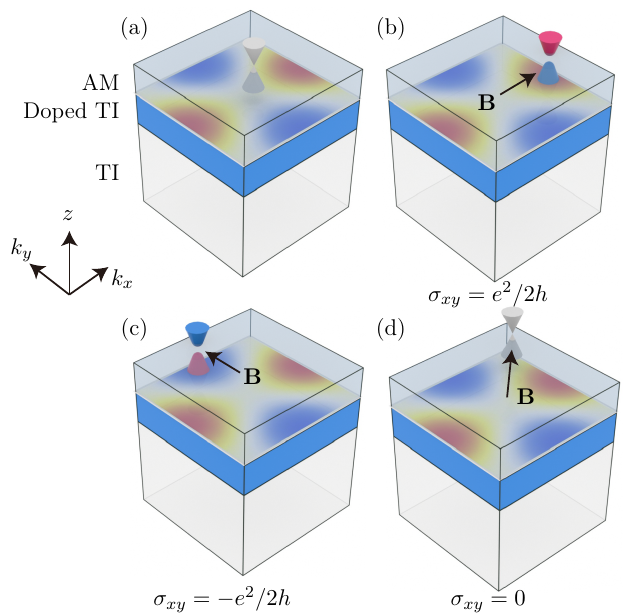}
\caption{Schematic illustration of the altermagnet/doped topological insulator/topological insulator heterostructure. The color scheme on the surface of the doped topological insulator indicates that the local magnetic moment originates from the altermagnet layer. (a) In the absence of the external in-plane magnetic field, the Dirac fermion on the topological insulator surface remains massless because there is no local magnetization at the Dirac point. (b)-(d) The Dirac point can be shifted by the external in-plane magnetic field. The shifted surface Dirac fermion gains a mass with its amplitude and sign being determined by the local magnetic moment. In experiments, the sign and amplitude of the Dirac mass can be extracted by measuring the Hall conductance. Thus, this method provides a probe for detecting the altermagnet texture.}
\vspace{-4em}
\label{fig:device}
\end{figure}

The interplay between topology and magnetism has led to the experimental discovery of various novel phases, such as the quantum anomalous Hall effect without an external magnetic field~\cite{Chang2013Science,Deng20sci}, the axion insulator with a topological magnetoelectric effect~\cite{Mogi17nm,Liu20nm}, and the semi-magnetic topological insulator with a half-quantized Hall conductance that provides evidence for the parity anomaly~\cite{mogi2021experimental}. These topological phases have been widely confirmed in  ferromagnetic topological insulator such as Cr-doped Bi$_2$Se$_3$ and antiferromagnetic topological insulators such as MnBi$_2$Te$_4$ ~\cite{Tokura19nrp,Chang2020NMRev,Wang2023}. These significant advancements naturally raise questions about the role of altermagnetism in topological insulators.

In this Letter, we propose to a method to detect the unique $\bm{k}$-space spin polarization of an altermagnets by placing them in proximity to the surface of a topological insulator~(Fig.~\ref{fig:device}). The surface of a topological insulator hosts a massless Dirac fermion with linear dispersion ~[Fig.~\ref{fig:device}(a)]~\cite{FL07PRL,Hasan2010RMP}. When time-reversal symmetry is broken by the adjacent altermagnet, the surface Dirac fermion gains a mass, with its sign and amplitude determined by the local magnetic moment at the Dirac point~[Figs.~\ref{fig:device}(b)-\ref{fig:device}(d)]. This massive Dirac fermion gives rise to an emergent half-quantized surface Hall conductance, allowing the local magnetic moment to be detected through Hall conductance measurements. Furthermore, the position of the Dirac point can be modulated by applying an in-plane magnetic field. These features allow us to use the relativistic Dirac fermion on the topological insulator surface as a probe to detect the unique nonrelativistic spin-momentum locking of altermagnets.

{\color{blue}\emph{Model}.}---We consider magnetic-doped topological insulator $\mathrm{Bi}_{2}\mathrm{Se}_{3}$ in proximity to altermagnetic layers~[Fig.~\ref{fig:device}] with the Hamiltonian
\begin{equation}
H=H_{\text{TI}}+H_{\Delta}+H_{J}.
\end{equation}
 $H_\text{TI}=\mathcal{M}\sigma_0 \tau_z+A_1 k_z \sigma_0 \tau_y
+A_2 ( k_x \sigma_x - k_y \sigma_y  )\tau_x$ describes topological insulator $\mathrm{Bi}_{2}\mathrm{Se}_{3}$ for the electrons of $P_{z,\uparrow}$ and $P_{z,\downarrow}$
orbitals from $\mathrm{Bi}$ and $\mathrm{Te}$ atoms near the Fermi energy, where $\mathcal{M}(\bm{k})=M_0-B_1k_z^2-
B_2(k_x^2+k_y^2)$~\cite{Liu10prb,Zhang09np}. $\sigma$ and $\tau$ are Pauli matrices representing spin and orbital degrees of freedom, respectively. We take $M=0.28$ eV, $A_1=0.22$ eV~nm, $A_2=0.41$ eV~nm, $B_1=0.1$ eV~nm$^2$, and $B_2=0.566$ eV~nm$^2$, which describes $\mathrm{Bi}_{2}\mathrm{Se}_{3}$~\cite{Zhang09np}. In the following calculations, we fix the film thickness as $L_z=10$ nm.

The second term $H_\Delta=F(z)\Delta(\cos\phi\sigma_x+\sin\phi\sigma_y)\tau_0$, represents the Zeeman type of spin splitting~\cite{Liu10prb}. We take $F(z)=1$ when $8 \text{ nm}<z<10\text{ nm}$, and $F(z)=0$ elsewhere, to indicate the magnetization effect on the top surface layers~\cite{mogi2021experimental}. $\Delta$ can originate from the direct Zeeman coupling between the electron spin and the external in-plane magnetic fields, or from the exchange coupling between the electron spin
and the magnetization of the magnetic ions~\cite{Liu2013}. The contribution from the former is weak (about $\Delta=0.001$ eV), but in magnetically doped systems, the exchange coupling is significantly enhanced. For example, the strength of exchange coupling around $\Delta=0.2$ eV can be introduced by about $5\%$ Cr-doping in Bi$_2$Se$_3$~\cite{Yu2010Science}, which is subjected to a weak external magnetic field about $B=0.2$ T~\cite{Chang2013Science,Kandala2015NC}. The angle $\phi$ represents the direction of the exchange coupling and can be adjusted by tuning the direction of the external in-plane magnetic field.

The third term represents the altermagnetic ordering achieved via the proximity effect with $H_J=F(z)J(k_x,k_y)\sigma_z \tau_0$, with $J(k_x,k_y)$ representing the form factor of altermagnetic ordering. Besides, the impact of the in-plane magnetic field on the altermagnet can be safely ignored. For example, the saturation of the magnetization in the altermagnets RuO$_2$~\cite{Tschirner2023APL,Feng2022NatElec} and MnTe~\cite{Gonzalez2023PRL} requires high magnetic fields of about 68 T and 4 T, respectively.

We derive the effective Hamiltonian for the surface states in an open boundary condition along the vertical direction (See Sec. SI of the Supplemental Material~\cite{Supp}), 
\begin{equation}
H^\prime=A_2 \left[(k_x-k_x^0)\sigma_x+(k_y-k_y^0)\sigma_y\right]
+J(k_x,k_y)\sigma_z.
\label{Eq:Eff}
\end{equation}
with $k_x^0=\Delta\cos(\phi)/A_2$ and $k_y^0=\Delta\sin(\phi)/A_2$.
In the absence of magnetic effects, i.e., $\Delta=J=0$, the surface state manifests a massless Dirac cone at the $\Gamma$
point~\cite{Shen2017TI}. When exchange coupling term $\Delta$ is introduced, the Dirac point shifts to $(k_x^0,k_y^0)$. This shifted Dirac fermion acquires a mass term proportional to the form factor of altermagnetic ordering $J(k_x^0,k_y^0)$. The Dirac mass can be detected by measuring the Hall conductance~\cite{Qi2011RMP}.
The Hall conductance of the massive Dirac fermion exhibits a half-quantized Hall conductance plateau when the Fermi energy resides in the magnetic gap~\cite{mogi2021experimental}. The width of the half-quantized Hall conductance plateau is proportional to the amplitude of $J(k_x^0,k_y^0)$, and the sign of the plateau is determined by the sign of $J(k_x^0,k_y^0)$. Specifically, in Sec.~SII of the Supplemental Material~\cite{Supp}, we show that the Hall conductance can be approximately expressed as
\begin{equation}
\sigma_{xy} \approx\left\{
\protect\begin{array}
[c]{ll}%
-\frac{e^2\text{sign}\left[J\left( k_x^0 ,k_y^0\right)
\right]}{2h}, &
\left\vert E_{F}\right\vert \leq\left\vert J\left( k_x^0 ,k_y^0\right)  \right\vert \protect,\\-\frac{e^2J\left( k_x^0 ,k_y^0\right)  }{2h\left|E_{F}\right|},& \left\vert E_{F}\right\vert >\left\vert J\left( k_x^0 ,k_y^0\right)  \right\vert
\protect.\end{array}
\right.
\label{Eq:Hall}
\end{equation}
The amplitude of $J(k_x^0,k_y^0)$ at a specific momentum-space point can be detected by measuring the Hall conductance for a given $\Delta$ and $\phi$. Then, adjusting the in-plane magnetic field to vary $\Delta$ and $\phi$ allows for mapping the distribution $J(k_x,k_y)$ in experiments.

\begin{figure}[t]
\centering
\includegraphics[width=\columnwidth]{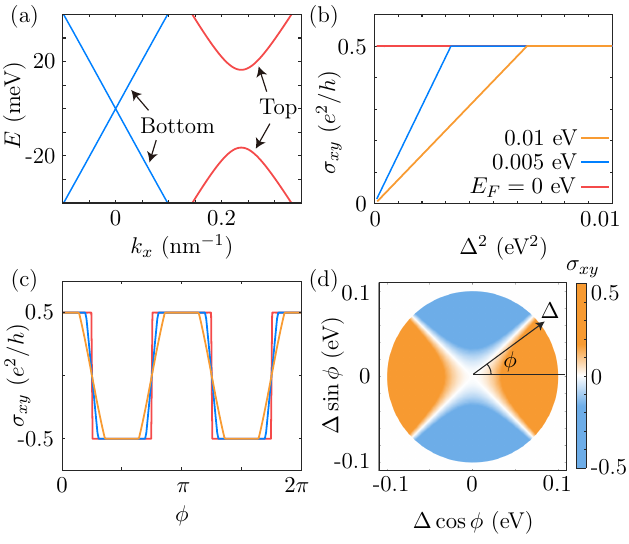}
\caption{Numerically calculated spectrum and Hall conductance for the $d$-wave altermagnetic system. (a) The bottom surface Dirac cone located at the $\Gamma$ point remains gapless, and the top surface Dirac cone shifts and opens a gap. (b) and (c) show the numerically calculated Hall conductance $\sigma_{xy}$ as a function of $\Delta^2$ and $\phi$, respectively. In (b)-(c), red, blue, and yellow correspond to different Fermi energies with $E_F=0$, $0.005$, and $0.01$ eV. (d) $\sigma_{xy}$ as functions of $\Delta$ and $\phi$. The parameters are $\Delta=0.1$ eV and $\phi=0$ in (a), $\phi=0$ in (b), $\Delta=0.1$ in (c), and $E_F=0.005$ eV in (d). Here, the results are calculated by using the tight-binding model in Sec.~SIII A of the Supplemental Material~\cite{Supp}.}
\label{fig:results}
\end{figure}

{\color{blue}\emph{Numerical results for $d$-wave altermagnets}.}---We consider a $d$-wave altermagnet with its form factor expressed as $J(k_x,k_y)=J_d(k_y^2-k_x^2)$. In the numerical calculations, we discretize the Hamiltonian on a 3D simple cubic lattice with lattice constant $a=1$ nm (Sec.~SIII A of the Supplemental Material~\cite{Supp}) and take the altermagnetization to be of the same order as the $k^2$ term in Bi$_2$Se$_3$ with $J_d = B_2$.

Figure~\ref{fig:results}(a) shows the energy spectrum of the system with $\phi=0$. The Dirac cone on the bottom surface remains massless because the magnetization vanishes on the bottom surface. The Dirac cone on the top surface shifts away from the $\Gamma$ point and gains a mass. The system is characterized by a half-quantized Hall conductance $\sigma_{xy}=e^2/2h$ when the Fermi energy $E_F$ is located inside the energy gap of the top surface Dirac cone~\cite{mogi2021experimental,Shen2017TI,Fu2022NPJQM}.

Figure~\ref{fig:results}(b) shows the Hall conductance as a function of $\Delta^2$ with $\phi=0$. For $E_F=0$, the Hall conductance remains the half-quantization as $\Delta^2$ varies. For a finite Fermi energy, i.e., $E_F\neq 0$, $\sigma_{xy}$ rises and reaches a half-quantized value. This scenario is explained as follows. When $\Delta^2$ is small, the Dirac cones is located near the $\Gamma$ point and gains a tiny topological mass due to the weak local magnetization, and thus $\sigma_{xy}$ deviates from the half-quantization because the Fermi energy crosses the energy bands. For a larger $\Delta^2$, the Dirac cone shifts away from $\Gamma$ point and opens a larger energy gap due to the stronger magnetization. Then $\sigma_{xy}=e^2/2h$ because the Fermi energy is located inside the energy gap.

In the regime where the conductance deviates from the half-quantization, the Hall conductance is proportional to $\Delta^2$. From Eq.~\eqref{Eq:Hall}, we have
\begin{equation}
\sigma_{xy} \approx\left\{
\protect\begin{array}
[c]{ll}%
\frac{e^2\text{sign}\left[\Delta^2J_d\cos 2\phi/A_2^2
\right]}{2h}, &
\left\vert E_{F}\right\vert \leq\left\vert \Delta^2J_d\cos 2\phi/A_2^2  \right\vert \protect,\\\frac{e^2\Delta^2J_d\cos 2\phi  }{2h\left|E_{F}\right|A_2^2},& \left\vert E_{F}\right\vert >\left\vert \Delta^2J_d\cos 2\phi/A_2^2  \right\vert
,\end{array}
\right.
\end{equation}
This explains the squared growth relationship between $\sigma_{xy}$ and $\Delta$. More importantly, this squared relationship originates from the fact that in the $d$-wave altermagnet, the magnetism near the $\Gamma$ point also varies quadratically with the momentum.

\begin{figure}[t]
\centering
\includegraphics[width=\columnwidth]{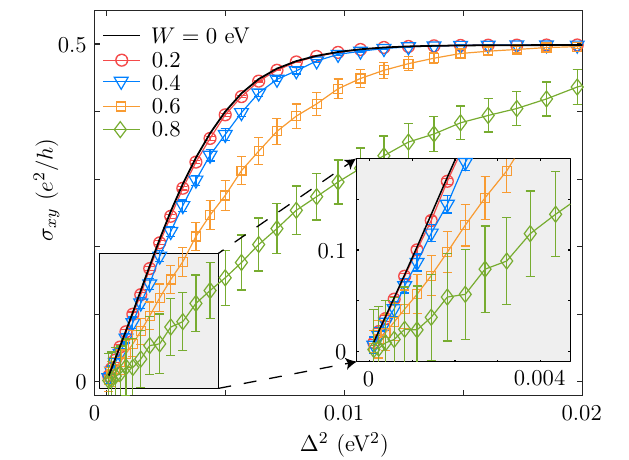}
\caption{Numerically calculated Hall conductance $\sigma_{xy}$ as a function of $\Delta^2$ with $E_F=0.005$ eV and different disorder strengths $W$. The error bars show the standard deviation for 1000 samples. The system size is $L_x=L_y=80$ nm and $L_z=10$ nm. Here, the results are calculated by using the tight-binding model in Sec.~SIII A of the Supplemental Material~\cite{Supp}.}
\label{fig:disorder}
\end{figure}

Now we show that more information can be obtained by tuning the angle $\phi$ [Fig.~\ref{fig:results}(c)]. For $\phi=\pi/2$, we have $\sigma_{xy}=-e^2/2h$ because the Dirac cone shifts to the $k_y$-direction~[Fig.~\ref{fig:device}(c)], acquiring an opposite mass compared to the case with $\phi=\pi/2$. For $\phi=\pi/4$, we have $\sigma_{xy}=0$ because the Dirac cone is shifted to the $k_x=k_y$ direction, and the magnetization vanishes along this direction [Fig.~\ref{fig:device}(d)]. This $\pi$-period of the Hall conductance arises from the $C_4 T$ symmetry of the $d$-wave altermagnetism, characterized by $\sigma_{xy}(\phi)=-\sigma_{xy}(\phi+\pi/2)$.

Figure~\ref{fig:results}(d) shows the numerically calculated Hall conductance in the $(\Delta\cos\phi,\Delta\sin\phi)$ plane, and we find the measured Hall conductance exhibits a pattern similar to the magnetic moment distribution of the $d$-wave altermagnet. This method gives a direct measurement for the $\bm{k}$-space spin density of altermagnets.

{\color{blue}\emph{Disorder effects}.}---Now we show that the emergent Hall conductance is robust against disorder, by calculating the conductance using the
Landauer-B\"uttiker formula
\cite{Landauer1970Philosophical,Buttiker1988PRB,
Fisher1981PRB} and the recursive Green's
function method \cite{Mackinnon1985Zeitschrift,Metalidis2005PRB}. We adopt the Anderson-type disorder
by considering random on-site energies fluctuating in the energy interval $[-W,W]$, where $W$ is the disorder strength. Figure~\ref{fig:disorder} shows the Hall conductance as a function of $\Delta^2$, with different disorder strengths (Sec.~SIII B of the Supplemental Material~\cite{Supp}). The result indicates that the linear behavior (see the insets in Fig.~\ref{fig:disorder}) is robust against weak disorder until the disorder strength $W$ exceeds $0.4$ eV, which is much larger compared to the magnetic gap (about $0.02$ eV). The robustness of the Hall conductance originates from the robust band topology of the surface Dirac cone in topological insulator. Therefore, the massive Dirac cone, as well as the emergent Hall conductance, are stable as long as the disorder strength is not strong enough to induce a topological phase transition.

\begin{figure}[t]
\centering
\includegraphics[width=\columnwidth]{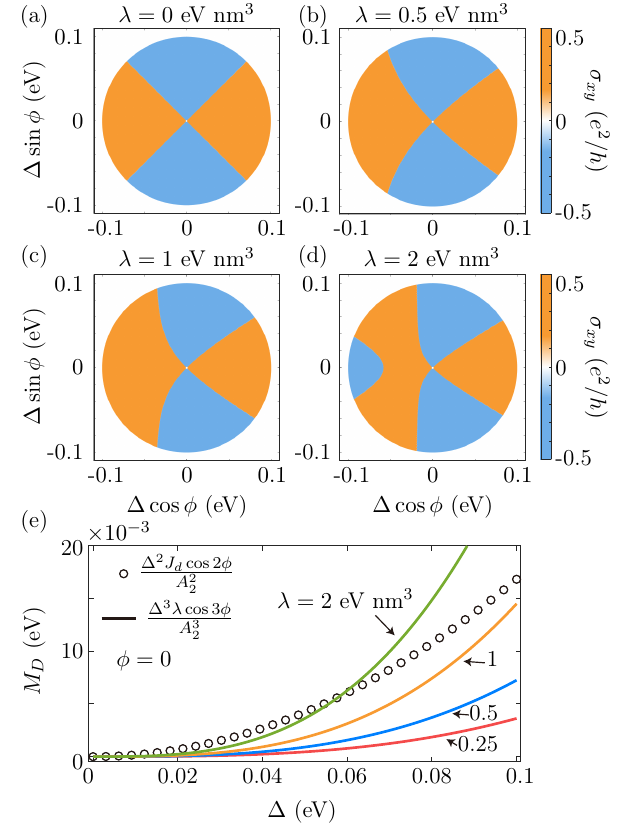}
\caption{(a)-(d) Hall conductance as functions of $\Delta$ and $\phi$ with different $\lambda$. (e) The circles and lines depict the contributions to the Dirac mass from the altermagnet and the warping effects, respectively. We take $E_F=0$ in all panels and $\phi=0$ in (e). Here, the results are calculated by using the effective Hamiltonian in Eq.~\eqref{Eq:Eff}.}
\label{fig:C6}
\end{figure}

{\color{blue}\emph{Hexagonal warping effect}.}--- It is worth noting that, the hexagonal warping effect in topological insulators may influence the measured Hall conductance by introducing a $2\pi/3$  period of the anomalous planar Hall effect~\cite{Liu2013}. However, in the following we will show that the transport of the system is still dominated by the altermagnetism for moderate strength of exchange coupling $\Delta$.

We include a warping term $H_\lambda=\lambda(k_x^3-3k_xk_y^2)\sigma_z$ to the effective Hamiltonian in Eq.~\eqref{Eq:Eff}. Figures~\ref{fig:C6}(a)-\ref{fig:C6}(d) show the numerical calculated Hall conductance as functions of $\Delta$ and $\phi$ for different $\lambda$ at $E_F=0$ eV. For $\lambda=0$, the Hall effect exhibits a $\pi$-period pattern as expected. For $\lambda=0.5$ and $1$ eV, the results show no significant difference compared to the case with $\lambda=0$, except for a slight modification at the phase boundary. Only when $\lambda$ reaches the significantly large value of around 2 eV and $\Delta$ is sufficiently large at approximately 0.1 eV, we observe that the Hall conductance exhibits a nearly $2\pi/3$ period as $\phi$ varies.

To better understand the above results, we analyze the contributions to the Dirac mass from the altermagnetization and warping effect, respectively. At the shifted Dirac point, the Dirac mass is determined by $M_D=\frac{\Delta^2 J_d \cos 2\phi}{A_2^2}+\frac{\Delta^3 \lambda \cos 3\phi}{A_2^3}$, where the former term arises from the altermagnetization and the latter term from the warping effect. For small $\Delta$, the Dirac mass is governed by the former squared term. This can be observed more clearly in Fig.~\ref{fig:C6}(e), we can see that the contribution from the altermagnetic term is significantly larger compared to that from the warping term, except in cases of extremely large
$\lambda$ about 2 eV.

In realistic materials, the warping term is about $\lambda=0.25$ eV~nm$^3$~\cite{FuL09PRL}. Therefore, the transport should be dominated by the altermagnism and the hexagonal warping effect can be ignored in experiments.

\begin{figure}[t]
\centering
\includegraphics[width=\columnwidth]{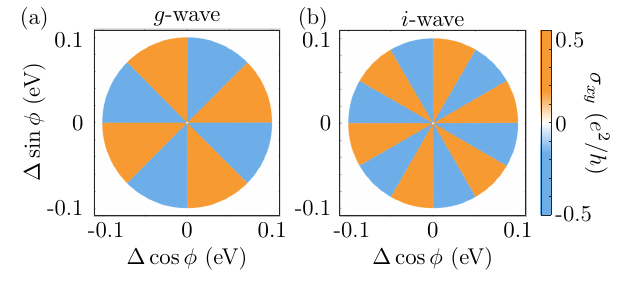}
\caption{Numerically calculated Hall conductance as functions of $\Delta$ and $\phi$ with (a) $g$-wave altermagnetism and (b) $i$-wave altermagnetism, respectively. We take $J_g=0.5$ eV nm$^4$ in (a) and $J_i=0.5$ eV nm$^6$ in (b). Here, the results are calculated by using the effective Hamiltonian in Eq.~\eqref{Eq:Eff}.}
\label{fig:giwave}
\end{figure}

\begin{table}[t]
\begin{ruledtabular}
\caption{Results for different altermagnets. The first column shows the Dirac mass of the heterostructure system for a certain $\Delta$ and $\phi$. The second column shows the period of the Hall conductance $\sigma_{xy}$ when tunes the angle $\phi$. The third column shows the dependence relationship between $\sigma_{xy}$ and $\Delta$ when the Fermi energy crosses the energy bands.}
\begin{tabular}{c|ccc}
&   $d$-wave&   $g$-wave&$i$-wave
\\\hline
$M_D$&$\frac{\Delta^2 J_d \cos 2\phi}{A_2^2} $&$\frac{\Delta^4 J_g \sin 4\phi}{4A_2^4}$&$\frac{\Delta^6 J_d \sin 6\phi}{2A_2^6}$
\\
period     &$\pi$&$\pi/2$
&$\pi/3$
\\
$\sigma_{xy}$&$\propto\Delta^2$&
$\propto\Delta^4$     &$\propto\Delta^6$
\\
\end{tabular}
\label{tab1}
\end{ruledtabular}
\end{table}

{\color{blue}\emph{$g$-wave and $i$-wave altermagnets}.}--- We propose that this approach can be used to detect altermagnets with $g$-wave or $i$-wave symmetry. Specifically, the $g$-wave altermagnet is characterized by $J(k_x,k_y)=J_gk_x k_y(k_x^2-k_y^2)$ and the $i$-wave altermagnet is described by $J(k_x,k_y)=J_ik_x k_y(3k_x^2-k_y^2)(k_x^2-3k_y^2)$, respectively~\cite{Smejkal22PRX}. The surface Dirac fermions gain different topological mass $M_D$ for a certain $\Delta$ and $\phi$~(the first column in Tab.~\ref{tab1}), leading to different periods of $\sigma_{xy}$ when $\phi$ is varied~(the second column in Tab.~\ref{tab1} and Fig.~\ref{fig:giwave}). Furthermore, when the Fermi energy crosses the energy bands, the Hall conductance exhibits different dependence relationship with the increasing $\Delta$~(the third column in Tab.~\ref{tab1}). These distinguishing features can be used to detect the unique magnetic texture of not only various altermagnets but also emerging materials with more complex magnetic textures.

Recent experimental work on Mn$_5$Si$_3$ shows that the relative Ne\'el vector orientation can be controlled by rotating the orientation of the external magnetic field. This change in the relative Ne\'el vector orientation yields anisotropic behavior of the anomalous Hall effect, which is different from that observed in ferromagnets. This suggests that epitaxial Mn$_5$Si$_3$ is closely match those expected for altermagnets~\cite{Leiviska2024PRB}. We believe that our methods can provide further evidence for detecting the altermagnets.

\begin{acknowledgments}
D.-H.X. was supported by the NSFC (under Grant Nos.~12474151, 12074108 and 12347101), the Natural Science Foundation of Chongqing (Grant No.~CSTB2022NSCQ-MSX0568) and Beijing National Laboratory for Condensed Matter Physics (No. 2024BNLCMPKF025). R.C. acknowledges the support of the NSFC (under Grant No. 12304195) and the Chutian Scholars Program in Hubei Province.  B.Z. was supported by the NSFC (under Grant No. 12074107), the program of outstanding young and middle-aged scientific and technological innovation team of colleges and universities in Hubei Province (under Grant No. T2020001) and the innovation group project of the natural science foundation of Hubei Province of China (under Grant No. 2022CFA012). H.P.S. was supported by the W\"urzburg-Dresden Cluster of Excellence ct.qmat (Project-id 390858490), the DFG (SFB 1170), and the Bavarian Ministry of Economic Affairs, Regional Development and Energy for financial support within the High-Tech Agenda Project ``Bausteine f\"ur das Quanten Computing auf Basis topologischer Materialen''.
\end{acknowledgments}

%
%
%
\bibliographystyle{apsrev4-1-etal-title_6authors}
\bibliography{refs-transport,refs-transport_v1}

\begin{thebibliography}{54}%
\makeatletter
\providecommand \@ifxundefined [1]{%
 \@ifx{#1\undefined}
}%
\providecommand \@ifnum [1]{%
 \ifnum #1\expandafter \@firstoftwo
 \else \expandafter \@secondoftwo
 \fi
}%
\providecommand \@ifx [1]{%
 \ifx #1\expandafter \@firstoftwo
 \else \expandafter \@secondoftwo
 \fi
}%
\providecommand \natexlab [1]{#1}%
\providecommand \enquote  [1]{``#1''}%
\providecommand \bibnamefont  [1]{#1}%
\providecommand \bibfnamefont [1]{#1}%
\providecommand \citenamefont [1]{#1}%
\providecommand \href@noop [0]{\@secondoftwo}%
\providecommand \href [0]{\begingroup \@sanitize@url \@href}%
\providecommand \@href[1]{\@@startlink{#1}\@@href}%
\providecommand \@@href[1]{\endgroup#1\@@endlink}%
\providecommand \@sanitize@url [0]{\catcode `\\12\catcode `\$12\catcode
  `\&12\catcode `\#12\catcode `\^12\catcode `\_12\catcode `\%12\relax}%
\providecommand \@@startlink[1]{}%
\providecommand \@@endlink[0]{}%
\providecommand \url  [0]{\begingroup\@sanitize@url \@url }%
\providecommand \@url [1]{\endgroup\@href {#1}{\urlprefix }}%
\providecommand \urlprefix  [0]{URL }%
\providecommand \Eprint [0]{\href }%
\providecommand \doibase [0]{http://dx.doi.org/}%
\providecommand \selectlanguage [0]{\@gobble}%
\providecommand \bibinfo  [0]{\@secondoftwo}%
\providecommand \bibfield  [0]{\@secondoftwo}%
\providecommand \translation [1]{[#1]}%
\providecommand \BibitemOpen [0]{}%
\providecommand \bibitemStop [0]{}%
\providecommand \bibitemNoStop [0]{.\EOS\space}%
\providecommand \EOS [0]{\spacefactor3000\relax}%
\providecommand \BibitemShut  [1]{\csname bibitem#1\endcsname}%
\let\auto@bib@innerbib\@empty
\bibitem [{\citenamefont {\ifmmode~\check{S}\else \v{S}\fi{}mejkal}\ \emph
  {et~al.}(2022{\natexlab{a}})\citenamefont {\ifmmode~\check{S}\else
  \v{S}\fi{}mejkal}, \citenamefont {Sinova},\ and\ \citenamefont
  {Jungwirth}}]{Smejkal22PRX}%
  \BibitemOpen
  \bibfield  {author} {\bibinfo {author} {\bibfnamefont {L.}~\bibnamefont
  {\ifmmode~\check{S}\else \v{S}\fi{}mejkal}}, \bibinfo {author} {\bibfnamefont
  {J.}~\bibnamefont {Sinova}}, \ and\ \bibinfo {author} {\bibfnamefont
  {T.}~\bibnamefont {Jungwirth}},\ }\bibfield  {title} {\enquote {\bibinfo
  {title} {Beyond conventional ferromagnetism and antiferromagnetism: A phase
  with nonrelativistic spin and crystal rotation symmetry}}, }\href {\doibase
  10.1103/PhysRevX.12.031042} {\bibfield  {journal} {\bibinfo  {journal} {Phys.
  Rev. X}\ }\textbf {\bibinfo {volume} {12}},\ \bibinfo {pages} {031042}
  (\bibinfo {year} {2022}{\natexlab{a}})}\BibitemShut {NoStop}%
\bibitem [{\citenamefont {Bai}\ \emph {et~al.}(2024)\citenamefont {Bai},
  \citenamefont {Feng}, \citenamefont {Liu}, \citenamefont {Šmejkal},
  \citenamefont {Mokrousov},\ and\ \citenamefont {Yao}}]{Bai2024AFM}%
  \BibitemOpen
  \bibfield  {author} {\bibinfo {author} {\bibfnamefont {L.}~\bibnamefont
  {Bai}}, \bibinfo {author} {\bibfnamefont {W.}~\bibnamefont {Feng}}, \bibinfo
  {author} {\bibfnamefont {S.}~\bibnamefont {Liu}}, \bibinfo {author}
  {\bibfnamefont {L.}~\bibnamefont {Šmejkal}}, \bibinfo {author}
  {\bibfnamefont {Y.}~\bibnamefont {Mokrousov}}, \ and\ \bibinfo {author}
  {\bibfnamefont {Y.}~\bibnamefont {Yao}},\ }\bibfield  {title} {\enquote
  {\bibinfo {title} {Altermagnetism: Exploring new frontiers in magnetism and
  spintronics}}, }\href {\doibase 10.1002/adfm.202409327} {\bibfield  {journal}
  {\bibinfo  {journal} {Adv. Funct. Mater.}\ ,\ \bibinfo {pages} {2409327}}
  (\bibinfo {year} {2024})}\BibitemShut {NoStop}%
\bibitem [{\citenamefont {{\v{S}}mejkal}\ \emph {et~al.}(2020)\citenamefont
  {{\v{S}}mejkal}, \citenamefont {Gonz{\'a}lez-Hern{\'a}ndez}, \citenamefont
  {Jungwirth},\ and\ \citenamefont {Sinova}}]{vsmejkal2020crystal}%
  \BibitemOpen
  \bibfield  {author} {\bibinfo {author} {\bibfnamefont {L.}~\bibnamefont
  {{\v{S}}mejkal}}, \bibinfo {author} {\bibfnamefont {R.}~\bibnamefont
  {Gonz{\'a}lez-Hern{\'a}ndez}}, \bibinfo {author} {\bibfnamefont
  {T.}~\bibnamefont {Jungwirth}}, \ and\ \bibinfo {author} {\bibfnamefont
  {J.}~\bibnamefont {Sinova}},\ }\bibfield  {title} {\enquote {\bibinfo {title}
  {{Crystal time-reversal symmetry breaking and spontaneous Hall effect in
  collinear antiferromagnets}}}, }\href
  {https://www.science.org/doi/full/10.1126/sciadv.aaz8809} {\bibfield
  {journal} {\bibinfo  {journal} {Sci. Adv.}\ }\textbf {\bibinfo {volume}
  {6}},\ \bibinfo {pages} {eaaz8809} (\bibinfo {year} {2020})}\BibitemShut
  {NoStop}%
\bibitem [{\citenamefont {Naka}\ \emph {et~al.}(2019)\citenamefont {Naka},
  \citenamefont {Hayami}, \citenamefont {Kusunose}, \citenamefont {Yanagi},
  \citenamefont {Motome},\ and\ \citenamefont {Seo}}]{Naka19NC}%
  \BibitemOpen
  \bibfield  {author} {\bibinfo {author} {\bibfnamefont {M.}~\bibnamefont
  {Naka}}, \bibinfo {author} {\bibfnamefont {S.}~\bibnamefont {Hayami}},
  \bibinfo {author} {\bibfnamefont {H.}~\bibnamefont {Kusunose}}, \bibinfo
  {author} {\bibfnamefont {Y.}~\bibnamefont {Yanagi}}, \bibinfo {author}
  {\bibfnamefont {Y.}~\bibnamefont {Motome}}, \ and\ \bibinfo {author}
  {\bibfnamefont {H.}~\bibnamefont {Seo}},\ }\bibfield  {title} {\enquote
  {\bibinfo {title} {Spin current generation in organic antiferromagnets}},
  }\href {\doibase 10.1038/s41467-019-12229-y} {\bibfield  {journal} {\bibinfo
  {journal} {Nat. Commun.}\ }\textbf {\bibinfo {volume} {10}},\ \bibinfo
  {pages} {4305} (\bibinfo {year} {2019})}\BibitemShut {NoStop}%
\bibitem [{\citenamefont {Ahn}\ \emph {et~al.}(2019)\citenamefont {Ahn},
  \citenamefont {Hariki}, \citenamefont {Lee},\ and\ \citenamefont
  {Kune\ifmmode~\check{s}\else \v{s}\fi{}}}]{Ahn19PRB}%
  \BibitemOpen
  \bibfield  {author} {\bibinfo {author} {\bibfnamefont {K.-H.}\ \bibnamefont
  {Ahn}}, \bibinfo {author} {\bibfnamefont {A.}~\bibnamefont {Hariki}},
  \bibinfo {author} {\bibfnamefont {K.-W.}\ \bibnamefont {Lee}}, \ and\
  \bibinfo {author} {\bibfnamefont {J.}~\bibnamefont
  {Kune\ifmmode~\check{s}\else \v{s}\fi{}}},\ }\bibfield  {title} {\enquote
  {\bibinfo {title} {{Antiferromagnetism in ${\mathrm{RuO}}_{2}$ as $d$-wave
  Pomeranchuk instability}}}, }\href {\doibase 10.1103/PhysRevB.99.184432}
  {\bibfield  {journal} {\bibinfo  {journal} {Phys. Rev. B}\ }\textbf {\bibinfo
  {volume} {99}},\ \bibinfo {pages} {184432} (\bibinfo {year}
  {2019})}\BibitemShut {NoStop}%
\bibitem [{\citenamefont {Hayami}\ \emph {et~al.}(2019)\citenamefont {Hayami},
  \citenamefont {Yanagi},\ and\ \citenamefont {Kusunose}}]{hayami2019momentum}%
  \BibitemOpen
  \bibfield  {author} {\bibinfo {author} {\bibfnamefont {S.}~\bibnamefont
  {Hayami}}, \bibinfo {author} {\bibfnamefont {Y.}~\bibnamefont {Yanagi}}, \
  and\ \bibinfo {author} {\bibfnamefont {H.}~\bibnamefont {Kusunose}},\
  }\bibfield  {title} {\enquote {\bibinfo {title} {Momentum-dependent spin
  splitting by collinear antiferromagnetic ordering}}, }\href
  {https://doi.org/10.7566/JPSJ.88.123702} {\bibfield  {journal} {\bibinfo
  {journal} {J. Phys. Soc. Jpn.}\ }\textbf {\bibinfo {volume} {88}},\ \bibinfo
  {pages} {123702} (\bibinfo {year} {2019})}\BibitemShut {NoStop}%
\bibitem [{\citenamefont {Yuan}\ \emph {et~al.}(2020)\citenamefont {Yuan},
  \citenamefont {Wang}, \citenamefont {Luo}, \citenamefont {Rashba},\ and\
  \citenamefont {Zunger}}]{Yuan2020Giant}%
  \BibitemOpen
  \bibfield  {author} {\bibinfo {author} {\bibfnamefont {L.-D.}\ \bibnamefont
  {Yuan}}, \bibinfo {author} {\bibfnamefont {Z.}~\bibnamefont {Wang}}, \bibinfo
  {author} {\bibfnamefont {J.-W.}\ \bibnamefont {Luo}}, \bibinfo {author}
  {\bibfnamefont {E.~I.}\ \bibnamefont {Rashba}}, \ and\ \bibinfo {author}
  {\bibfnamefont {A.}~\bibnamefont {Zunger}},\ }\bibfield  {title} {\enquote
  {\bibinfo {title} {Giant momentum-dependent spin splitting in centrosymmetric
  low-$z$ antiferromagnets}}, }\href {\doibase 10.1103/PhysRevB.102.014422}
  {\bibfield  {journal} {\bibinfo  {journal} {Phys. Rev. B}\ }\textbf {\bibinfo
  {volume} {102}},\ \bibinfo {pages} {014422} (\bibinfo {year}
  {2020})}\BibitemShut {NoStop}%
\bibitem [{\citenamefont {Mazin}\ \emph {et~al.}(2021)\citenamefont {Mazin},
  \citenamefont {Koepernik}, \citenamefont {Johannes}, \citenamefont
  {Gonz{\'a}lez-Hern{\'a}ndez},\ and\ \citenamefont
  {{\v{S}}mejkal}}]{mazin2021prediction}%
  \BibitemOpen
  \bibfield  {author} {\bibinfo {author} {\bibfnamefont {I.~I.}\ \bibnamefont
  {Mazin}}, \bibinfo {author} {\bibfnamefont {K.}~\bibnamefont {Koepernik}},
  \bibinfo {author} {\bibfnamefont {M.~D.}\ \bibnamefont {Johannes}}, \bibinfo
  {author} {\bibfnamefont {R.}~\bibnamefont {Gonz{\'a}lez-Hern{\'a}ndez}}, \
  and\ \bibinfo {author} {\bibfnamefont {L.}~\bibnamefont {{\v{S}}mejkal}},\
  }\bibfield  {title} {\enquote {\bibinfo {title} {{Prediction of
  unconventional magnetism in doped FeSb$_2$}}}, }\href
  {http://dx.doi.org/10.1073/pnas.2108924118} {\bibfield  {journal} {\bibinfo
  {journal} {Proc. Nat. Acad. Sci.}\ }\textbf {\bibinfo {volume} {118}},\
  \bibinfo {pages} {e2108924118} (\bibinfo {year} {2021})}\BibitemShut
  {NoStop}%
\bibitem [{\citenamefont {Ma}\ \emph {et~al.}(2021)\citenamefont {Ma},
  \citenamefont {Hu}, \citenamefont {Li}, \citenamefont {Liu}, \citenamefont
  {Yao}, \citenamefont {Jia},\ and\ \citenamefont
  {Liu}}]{ma2021multifunctional}%
  \BibitemOpen
  \bibfield  {author} {\bibinfo {author} {\bibfnamefont {H.-Y.}\ \bibnamefont
  {Ma}}, \bibinfo {author} {\bibfnamefont {M.}~\bibnamefont {Hu}}, \bibinfo
  {author} {\bibfnamefont {N.}~\bibnamefont {Li}}, \bibinfo {author}
  {\bibfnamefont {J.}~\bibnamefont {Liu}}, \bibinfo {author} {\bibfnamefont
  {W.}~\bibnamefont {Yao}}, \bibinfo {author} {\bibfnamefont {J.-F.}\
  \bibnamefont {Jia}}, \ and\ \bibinfo {author} {\bibfnamefont
  {J.}~\bibnamefont {Liu}},\ }\bibfield  {title} {\enquote {\bibinfo {title}
  {Multifunctional antiferromagnetic materials with giant piezomagnetism and
  noncollinear spin current}}, }\href
  {https://doi.org/10.1038/s41467-021-23127-7} {\bibfield  {journal} {\bibinfo
  {journal} {Nat. Commun.}\ }\textbf {\bibinfo {volume} {12}},\ \bibinfo
  {pages} {2846} (\bibinfo {year} {2021})}\BibitemShut {NoStop}%
\bibitem [{\citenamefont {\ifmmode~\check{S}\else \v{S}\fi{}mejkal}\ \emph
  {et~al.}(2022{\natexlab{b}})\citenamefont {\ifmmode~\check{S}\else
  \v{S}\fi{}mejkal}, \citenamefont {Sinova},\ and\ \citenamefont
  {Jungwirth}}]{ifmmode2022Emerging}%
  \BibitemOpen
  \bibfield  {author} {\bibinfo {author} {\bibfnamefont {L.}~\bibnamefont
  {\ifmmode~\check{S}\else \v{S}\fi{}mejkal}}, \bibinfo {author} {\bibfnamefont
  {J.}~\bibnamefont {Sinova}}, \ and\ \bibinfo {author} {\bibfnamefont
  {T.}~\bibnamefont {Jungwirth}},\ }\bibfield  {title} {\enquote {\bibinfo
  {title} {Emerging research landscape of altermagnetism}}, }\href {\doibase
  10.1103/PhysRevX.12.040501} {\bibfield  {journal} {\bibinfo  {journal} {Phys.
  Rev. X}\ }\textbf {\bibinfo {volume} {12}},\ \bibinfo {pages} {040501}
  (\bibinfo {year} {2022}{\natexlab{b}})}\BibitemShut {NoStop}%
\bibitem [{\citenamefont {Krempask{\`y}}\ \emph {et~al.}(2024)\citenamefont
  {Krempask{\`y}}, \citenamefont {{\v{S}}mejkal}, \citenamefont {D’souza},
  \citenamefont {Hajlaoui}, \citenamefont {Springholz}, \citenamefont
  {Uhl{\'\i}{\v{r}}ov{\'a}}, \citenamefont {Alarab}, \citenamefont
  {Constantinou}, \citenamefont {Strocov}, \citenamefont {Usanov} \emph
  {et~al.}}]{krempasky2024altermagnetic}%
  \BibitemOpen
  \bibfield  {author} {\bibinfo {author} {\bibfnamefont {J.}~\bibnamefont
  {Krempask{\`y}}}, \bibinfo {author} {\bibfnamefont {L.}~\bibnamefont
  {{\v{S}}mejkal}}, \bibinfo {author} {\bibfnamefont {S.}~\bibnamefont
  {D’souza}}, \bibinfo {author} {\bibfnamefont {M.}~\bibnamefont {Hajlaoui}},
  \bibinfo {author} {\bibfnamefont {G.}~\bibnamefont {Springholz}}, \bibinfo
  {author} {\bibfnamefont {K.}~\bibnamefont {Uhl{\'\i}{\v{r}}ov{\'a}}},  \emph
  {et~al.},\ }\bibfield  {title} {\enquote {\bibinfo {title} {Altermagnetic
  lifting of kramers spin degeneracy}}, }\href
  {https://doi.org/10.1038/s41586-023-06907-7} {\bibfield  {journal} {\bibinfo
  {journal} {Nature}\ }\textbf {\bibinfo {volume} {626}},\ \bibinfo {pages}
  {517} (\bibinfo {year} {2024})}\BibitemShut {NoStop}%
\bibitem [{\citenamefont {Li}\ \emph {et~al.}(2024)\citenamefont {Li},
  \citenamefont {Wang}, \citenamefont {Ding}, \citenamefont {Ren},
  \citenamefont {Zhao}, \citenamefont {Lin}, \citenamefont {Yang},
  \citenamefont {Yan}, \citenamefont {Li}, \citenamefont {Yang}, \citenamefont
  {Yuan}, \citenamefont {Denlinger}, \citenamefont {Wang}, \citenamefont
  {Zhang}, \citenamefont {Wray}, \citenamefont {Dong}, \citenamefont {Qian},\
  and\ \citenamefont {Miao}}]{sdongFeSb2}%
  \BibitemOpen
  \bibfield  {author} {\bibinfo {author} {\bibfnamefont {H.}~\bibnamefont
  {Li}}, \bibinfo {author} {\bibfnamefont {G.}~\bibnamefont {Wang}}, \bibinfo
  {author} {\bibfnamefont {N.}~\bibnamefont {Ding}}, \bibinfo {author}
  {\bibfnamefont {Q.}~\bibnamefont {Ren}}, \bibinfo {author} {\bibfnamefont
  {G.}~\bibnamefont {Zhao}}, \bibinfo {author} {\bibfnamefont {W.}~\bibnamefont
  {Lin}},  \emph {et~al.},\ }\bibfield  {title} {\enquote {\bibinfo {title}
  {{Spectroscopic evidence of spin-state excitation in d-electron correlated
  semiconductor FeSb$_2$}}}, }\href {https://doi.org/10.1073/pnas.2321193121}
  {\bibfield  {journal} {\bibinfo  {journal} {Proc. Nat. Acad. Sci.}\ }\textbf
  {\bibinfo {volume} {121}},\ \bibinfo {pages} {e2321193121} (\bibinfo {year}
  {2024})}\BibitemShut {NoStop}%
\bibitem [{\citenamefont {Naka}\ \emph {et~al.}(2021)\citenamefont {Naka},
  \citenamefont {Motome},\ and\ \citenamefont {Seo}}]{Nakaprb}%
  \BibitemOpen
  \bibfield  {author} {\bibinfo {author} {\bibfnamefont {M.}~\bibnamefont
  {Naka}}, \bibinfo {author} {\bibfnamefont {Y.}~\bibnamefont {Motome}}, \ and\
  \bibinfo {author} {\bibfnamefont {H.}~\bibnamefont {Seo}},\ }\bibfield
  {title} {\enquote {\bibinfo {title} {Perovskite as a spin current
  generator}}, }\href {\doibase 10.1103/PhysRevB.103.125114} {\bibfield
  {journal} {\bibinfo  {journal} {Phys. Rev. B}\ }\textbf {\bibinfo {volume}
  {103}},\ \bibinfo {pages} {125114} (\bibinfo {year} {2021})}\BibitemShut
  {NoStop}%
\bibitem [{\citenamefont {Shao}\ \emph {et~al.}(2021)\citenamefont {Shao},
  \citenamefont {Zhang}, \citenamefont {Li}, \citenamefont {Eom},\ and\
  \citenamefont {Tsymbal}}]{shao2021spin}%
  \BibitemOpen
  \bibfield  {author} {\bibinfo {author} {\bibfnamefont {D.-F.}\ \bibnamefont
  {Shao}}, \bibinfo {author} {\bibfnamefont {S.-H.}\ \bibnamefont {Zhang}},
  \bibinfo {author} {\bibfnamefont {M.}~\bibnamefont {Li}}, \bibinfo {author}
  {\bibfnamefont {C.-B.}\ \bibnamefont {Eom}}, \ and\ \bibinfo {author}
  {\bibfnamefont {E.~Y.}\ \bibnamefont {Tsymbal}},\ }\bibfield  {title}
  {\enquote {\bibinfo {title} {Spin-neutral currents for spintronics}}, }\href
  {https://doi.org/10.1038/s41467-021-26915-3} {\bibfield  {journal} {\bibinfo
  {journal} {Nat. Commun.}\ }\textbf {\bibinfo {volume} {12}},\ \bibinfo
  {pages} {7061} (\bibinfo {year} {2021})}\BibitemShut {NoStop}%
\bibitem [{\citenamefont {Feng}\ \emph
  {et~al.}(2022{\natexlab{a}})\citenamefont {Feng}, \citenamefont {Zhou},
  \citenamefont {{\v{S}}mejkal}, \citenamefont {Wu}, \citenamefont {Zhu},
  \citenamefont {Guo}, \citenamefont {Gonz{\'a}lez-Hern{\'a}ndez},
  \citenamefont {Wang}, \citenamefont {Yan}, \citenamefont {Qin} \emph
  {et~al.}}]{FengZX22NE}%
  \BibitemOpen
  \bibfield  {author} {\bibinfo {author} {\bibfnamefont {Z.}~\bibnamefont
  {Feng}}, \bibinfo {author} {\bibfnamefont {X.}~\bibnamefont {Zhou}}, \bibinfo
  {author} {\bibfnamefont {L.}~\bibnamefont {{\v{S}}mejkal}}, \bibinfo {author}
  {\bibfnamefont {L.}~\bibnamefont {Wu}}, \bibinfo {author} {\bibfnamefont
  {Z.}~\bibnamefont {Zhu}}, \bibinfo {author} {\bibfnamefont {H.}~\bibnamefont
  {Guo}},  \emph {et~al.},\ }\bibfield  {title} {\enquote {\bibinfo {title}
  {{An anomalous Hall effect in altermagnetic ruthenium dioxide}}}, }\href
  {https://www.nature.com/articles/s41928-022-00866-z} {\bibfield  {journal}
  {\bibinfo  {journal} {Nat. Electron.}\ }\textbf {\bibinfo {volume} {5}},\
  \bibinfo {pages} {735} (\bibinfo {year} {2022}{\natexlab{a}})}\BibitemShut
  {NoStop}%
\bibitem [{\citenamefont {\ifmmode~\check{S}\else \v{S}\fi{}mejkal}\ \emph
  {et~al.}(2022{\natexlab{c}})\citenamefont {\ifmmode~\check{S}\else
  \v{S}\fi{}mejkal}, \citenamefont {Hellenes}, \citenamefont
  {Gonz\'alez-Hern\'andez}, \citenamefont {Sinova},\ and\ \citenamefont
  {Jungwirth}}]{ifmmode2022Giant}%
  \BibitemOpen
  \bibfield  {author} {\bibinfo {author} {\bibfnamefont {L.}~\bibnamefont
  {\ifmmode~\check{S}\else \v{S}\fi{}mejkal}}, \bibinfo {author} {\bibfnamefont
  {A.~B.}\ \bibnamefont {Hellenes}}, \bibinfo {author} {\bibfnamefont
  {R.}~\bibnamefont {Gonz\'alez-Hern\'andez}}, \bibinfo {author} {\bibfnamefont
  {J.}~\bibnamefont {Sinova}}, \ and\ \bibinfo {author} {\bibfnamefont
  {T.}~\bibnamefont {Jungwirth}},\ }\bibfield  {title} {\enquote {\bibinfo
  {title} {Giant and tunneling magnetoresistance in unconventional collinear
  antiferromagnets with nonrelativistic spin-momentum coupling}}, }\href
  {\doibase 10.1103/PhysRevX.12.011028} {\bibfield  {journal} {\bibinfo
  {journal} {Phys. Rev. X}\ }\textbf {\bibinfo {volume} {12}},\ \bibinfo
  {pages} {011028} (\bibinfo {year} {2022}{\natexlab{c}})}\BibitemShut
  {NoStop}%
\bibitem [{\citenamefont {Fernandes}\ \emph {et~al.}(2024)\citenamefont
  {Fernandes}, \citenamefont {de~Carvalho}, \citenamefont {Birol},\ and\
  \citenamefont {Pereira}}]{Fernandestoplogical}%
  \BibitemOpen
  \bibfield  {author} {\bibinfo {author} {\bibfnamefont {R.~M.}\ \bibnamefont
  {Fernandes}}, \bibinfo {author} {\bibfnamefont {V.~S.}\ \bibnamefont
  {de~Carvalho}}, \bibinfo {author} {\bibfnamefont {T.}~\bibnamefont {Birol}},
  \ and\ \bibinfo {author} {\bibfnamefont {R.~G.}\ \bibnamefont {Pereira}},\
  }\bibfield  {title} {\enquote {\bibinfo {title} {Topological transition from
  nodal to nodeless zeeman splitting in altermagnets}}, }\href {\doibase
  10.1103/PhysRevB.109.024404} {\bibfield  {journal} {\bibinfo  {journal}
  {Phys. Rev. B}\ }\textbf {\bibinfo {volume} {109}},\ \bibinfo {pages}
  {024404} (\bibinfo {year} {2024})}\BibitemShut {NoStop}%
\bibitem [{\citenamefont {Zhang}\ \emph {et~al.}(2024)\citenamefont {Zhang},
  \citenamefont {Cui}, \citenamefont {Li}, \citenamefont {Duan}, \citenamefont
  {Li}, \citenamefont {Yu},\ and\ \citenamefont {Yao}}]{zhang2024prl}%
  \BibitemOpen
  \bibfield  {author} {\bibinfo {author} {\bibfnamefont {R.-W.}\ \bibnamefont
  {Zhang}}, \bibinfo {author} {\bibfnamefont {C.}~\bibnamefont {Cui}}, \bibinfo
  {author} {\bibfnamefont {R.}~\bibnamefont {Li}}, \bibinfo {author}
  {\bibfnamefont {J.}~\bibnamefont {Duan}}, \bibinfo {author} {\bibfnamefont
  {L.}~\bibnamefont {Li}}, \bibinfo {author} {\bibfnamefont {Z.-M.}\
  \bibnamefont {Yu}}, \ and\ \bibinfo {author} {\bibfnamefont {Y.}~\bibnamefont
  {Yao}},\ }\bibfield  {title} {\enquote {\bibinfo {title} {Predictable
  gate-field control of spin in altermagnets with spin-layer coupling}}, }\href
  {\doibase 10.1103/PhysRevLett.133.056401} {\bibfield  {journal} {\bibinfo
  {journal} {Phys. Rev. Lett.}\ }\textbf {\bibinfo {volume} {133}},\ \bibinfo
  {pages} {056401} (\bibinfo {year} {2024})}\BibitemShut {NoStop}%
\bibitem [{\citenamefont {Ke{\ss}ler}\ \emph {et~al.}(2024)\citenamefont
  {Ke{\ss}ler}, \citenamefont {Garcia-Gassull}, \citenamefont {Suter},
  \citenamefont {Prokscha}, \citenamefont {Salman}, \citenamefont {Khalyavin},
  \citenamefont {Manuel}, \citenamefont {Orlandi}, \citenamefont {Mazin},
  \citenamefont {Valentí},\ and\ \citenamefont {Moser}}]{Keler2024NpjS}%
  \BibitemOpen
  \bibfield  {author} {\bibinfo {author} {\bibfnamefont {P.}~\bibnamefont
  {Ke{\ss}ler}}, \bibinfo {author} {\bibfnamefont {L.}~\bibnamefont
  {Garcia-Gassull}}, \bibinfo {author} {\bibfnamefont {A.}~\bibnamefont
  {Suter}}, \bibinfo {author} {\bibfnamefont {T.}~\bibnamefont {Prokscha}},
  \bibinfo {author} {\bibfnamefont {Z.}~\bibnamefont {Salman}}, \bibinfo
  {author} {\bibfnamefont {D.}~\bibnamefont {Khalyavin}},  \emph {et~al.},\
  }\bibfield  {title} {\enquote {\bibinfo {title} {{Absence of magnetic order
  in RuO$_2$: insights from $\mu$SR spectroscopy and neutron diffraction}}},
  }\href {\doibase 10.1038/s44306-024-00055-y} {\bibfield  {journal} {\bibinfo
  {journal} {npj Spintronics}\ }\textbf {\bibinfo {volume} {2}},\ \bibinfo
  {pages} {50} (\bibinfo {year} {2024})}\BibitemShut {NoStop}%
\bibitem [{\citenamefont {Feng}\ \emph
  {et~al.}(2022{\natexlab{b}})\citenamefont {Feng}, \citenamefont {Zhou},
  \citenamefont {Smejkal}, \citenamefont {Wu}, \citenamefont {Zhu},
  \citenamefont {Guo}, \citenamefont {González-Hernández}, \citenamefont
  {Wang}, \citenamefont {Yan}, \citenamefont {Qin}, \citenamefont {Zhang},
  \citenamefont {Wu}, \citenamefont {Chen}, \citenamefont {Meng}, \citenamefont
  {Liu}, \citenamefont {Xia}, \citenamefont {Sinova}, \citenamefont
  {Jungwirth},\ and\ \citenamefont {Liu}}]{Feng2022NatElec}%
  \BibitemOpen
  \bibfield  {author} {\bibinfo {author} {\bibfnamefont {Z.}~\bibnamefont
  {Feng}}, \bibinfo {author} {\bibfnamefont {X.}~\bibnamefont {Zhou}}, \bibinfo
  {author} {\bibfnamefont {L.}~\bibnamefont {Smejkal}}, \bibinfo {author}
  {\bibfnamefont {L.}~\bibnamefont {Wu}}, \bibinfo {author} {\bibfnamefont
  {Z.}~\bibnamefont {Zhu}}, \bibinfo {author} {\bibfnamefont {H.}~\bibnamefont
  {Guo}},  \emph {et~al.},\ }\bibfield  {title} {\enquote {\bibinfo {title} {An
  anomalous {Hall} effect in altermagnetic ruthenium dioxide}}, }\href
  {\doibase 10.1038/s41928-022-00866-z} {\bibfield  {journal} {\bibinfo
  {journal} {Nat. Elec.}\ }\textbf {\bibinfo {volume} {5}},\ \bibinfo {pages}
  {735} (\bibinfo {year} {2022}{\natexlab{b}})}\BibitemShut {NoStop}%
\bibitem [{\citenamefont {Zhou}\ \emph {et~al.}(2024)\citenamefont {Zhou},
  \citenamefont {Feng}, \citenamefont {Zhang}, \citenamefont
  {\ifmmode~\check{S}\else \v{S}\fi{}mejkal}, \citenamefont {Sinova},
  \citenamefont {Mokrousov},\ and\ \citenamefont {Yao}}]{Zhou2024PRL}%
  \BibitemOpen
  \bibfield  {author} {\bibinfo {author} {\bibfnamefont {X.}~\bibnamefont
  {Zhou}}, \bibinfo {author} {\bibfnamefont {W.}~\bibnamefont {Feng}}, \bibinfo
  {author} {\bibfnamefont {R.-W.}\ \bibnamefont {Zhang}}, \bibinfo {author}
  {\bibfnamefont {L.}~\bibnamefont {\ifmmode~\check{S}\else \v{S}\fi{}mejkal}},
  \bibinfo {author} {\bibfnamefont {J.}~\bibnamefont {Sinova}}, \bibinfo
  {author} {\bibfnamefont {Y.}~\bibnamefont {Mokrousov}}, \ and\ \bibinfo
  {author} {\bibfnamefont {Y.}~\bibnamefont {Yao}},\ }\bibfield  {title}
  {\enquote {\bibinfo {title} {{Crystal Thermal Transport in Altermagnetic
  ${\mathrm{RuO}}_{2}$}}}, }\href {\doibase 10.1103/PhysRevLett.132.056701}
  {\bibfield  {journal} {\bibinfo  {journal} {Phys. Rev. Lett.}\ }\textbf
  {\bibinfo {volume} {132}},\ \bibinfo {pages} {056701} (\bibinfo {year}
  {2024})}\BibitemShut {NoStop}%
\bibitem [{\citenamefont {Karube}\ \emph {et~al.}(2022)\citenamefont {Karube},
  \citenamefont {Tanaka}, \citenamefont {Sugawara}, \citenamefont {Kadoguchi},
  \citenamefont {Kohda},\ and\ \citenamefont {Nitta}}]{Karube2022PRL}%
  \BibitemOpen
  \bibfield  {author} {\bibinfo {author} {\bibfnamefont {S.}~\bibnamefont
  {Karube}}, \bibinfo {author} {\bibfnamefont {T.}~\bibnamefont {Tanaka}},
  \bibinfo {author} {\bibfnamefont {D.}~\bibnamefont {Sugawara}}, \bibinfo
  {author} {\bibfnamefont {N.}~\bibnamefont {Kadoguchi}}, \bibinfo {author}
  {\bibfnamefont {M.}~\bibnamefont {Kohda}}, \ and\ \bibinfo {author}
  {\bibfnamefont {J.}~\bibnamefont {Nitta}},\ }\bibfield  {title} {\enquote
  {\bibinfo {title} {{Observation of Spin-Splitter Torque in Collinear
  Antiferromagnetic ${\mathrm{RuO}}_{2}$}}}, }\href {\doibase
  10.1103/PhysRevLett.129.137201} {\bibfield  {journal} {\bibinfo  {journal}
  {Phys. Rev. Lett.}\ }\textbf {\bibinfo {volume} {129}},\ \bibinfo {pages}
  {137201} (\bibinfo {year} {2022})}\BibitemShut {NoStop}%
\bibitem [{\citenamefont {Fedchenko}\ \emph {et~al.}(2024)\citenamefont
  {Fedchenko}, \citenamefont {Minar}, \citenamefont {Akashdeep}, \citenamefont
  {DSouza}, \citenamefont {Vasilyev}, \citenamefont {Tkach}, \citenamefont
  {Odenbreit}, \citenamefont {Nguyen}, \citenamefont {Kutnyakhov},
  \citenamefont {Wind}, \citenamefont {Wenthaus}, \citenamefont {Scholz},
  \citenamefont {Rossnagel}, \citenamefont {Hoesch}, \citenamefont
  {Aeschlimann}, \citenamefont {Stadtmüller}, \citenamefont {Klaui},
  \citenamefont {Schönhense}, \citenamefont {Jungwirth}, \citenamefont
  {Hellenes}, \citenamefont {Jakob}, \citenamefont {Smejkal}, \citenamefont
  {Sinova},\ and\ \citenamefont {Elmers}}]{Fedchenko2024SciAdv}%
  \BibitemOpen
  \bibfield  {author} {\bibinfo {author} {\bibfnamefont {O.}~\bibnamefont
  {Fedchenko}}, \bibinfo {author} {\bibfnamefont {J.}~\bibnamefont {Minar}},
  \bibinfo {author} {\bibfnamefont {A.}~\bibnamefont {Akashdeep}}, \bibinfo
  {author} {\bibfnamefont {S.~W.}\ \bibnamefont {DSouza}}, \bibinfo {author}
  {\bibfnamefont {D.}~\bibnamefont {Vasilyev}}, \bibinfo {author}
  {\bibfnamefont {O.}~\bibnamefont {Tkach}},  \emph {et~al.},\ }\bibfield
  {title} {\enquote {\bibinfo {title} {{Observation of time-reversal symmetry
  breaking in the band structure of altermagnetic RuO$_2$}}}, }\href {\doibase
  10.1126/sciadv.adj4883} {\bibfield  {journal} {\bibinfo  {journal} {Sci.
  Adv.}\ }\textbf {\bibinfo {volume} {10}},\ \bibinfo {pages} {eadj4883}
  (\bibinfo {year} {2024})}\BibitemShut {NoStop}%
\bibitem [{\citenamefont {Hiraishi}\ \emph {et~al.}(2024)\citenamefont
  {Hiraishi}, \citenamefont {Okabe}, \citenamefont {Koda}, \citenamefont
  {Kadono}, \citenamefont {Muroi}, \citenamefont {Hirai},\ and\ \citenamefont
  {Hiroi}}]{Hiraishi2024PRL}%
  \BibitemOpen
  \bibfield  {author} {\bibinfo {author} {\bibfnamefont {M.}~\bibnamefont
  {Hiraishi}}, \bibinfo {author} {\bibfnamefont {H.}~\bibnamefont {Okabe}},
  \bibinfo {author} {\bibfnamefont {A.}~\bibnamefont {Koda}}, \bibinfo {author}
  {\bibfnamefont {R.}~\bibnamefont {Kadono}}, \bibinfo {author} {\bibfnamefont
  {T.}~\bibnamefont {Muroi}}, \bibinfo {author} {\bibfnamefont
  {D.}~\bibnamefont {Hirai}}, \ and\ \bibinfo {author} {\bibfnamefont
  {Z.}~\bibnamefont {Hiroi}},\ }\bibfield  {title} {\enquote {\bibinfo {title}
  {{Nonmagnetic Ground State in ${\mathrm{RuO}}_{2}$ Revealed by Muon Spin
  Rotation}}}, }\href {\doibase 10.1103/PhysRevLett.132.166702} {\bibfield
  {journal} {\bibinfo  {journal} {Phys. Rev. Lett.}\ }\textbf {\bibinfo
  {volume} {132}},\ \bibinfo {pages} {166702} (\bibinfo {year}
  {2024})}\BibitemShut {NoStop}%
\bibitem [{\citenamefont {Liu}\ \emph {et~al.}(2024)\citenamefont {Liu},
  \citenamefont {Zhan}, \citenamefont {Li}, \citenamefont {Liu}, \citenamefont
  {Cheng}, \citenamefont {Shi}, \citenamefont {Deng}, \citenamefont {Zhang},
  \citenamefont {Li}, \citenamefont {Ding} \emph {et~al.}}]{liu2024arXiv}%
  \BibitemOpen
  \bibfield  {author} {\bibinfo {author} {\bibfnamefont {J.}~\bibnamefont
  {Liu}}, \bibinfo {author} {\bibfnamefont {J.}~\bibnamefont {Zhan}}, \bibinfo
  {author} {\bibfnamefont {T.}~\bibnamefont {Li}}, \bibinfo {author}
  {\bibfnamefont {J.}~\bibnamefont {Liu}}, \bibinfo {author} {\bibfnamefont
  {S.}~\bibnamefont {Cheng}}, \bibinfo {author} {\bibfnamefont
  {Y.}~\bibnamefont {Shi}},  \emph {et~al.},\ }\bibfield  {title} {\enquote
  {\bibinfo {title} {{Absence of altermagnetic spin splitting character in
  rutile oxide RuO$_2$}}}, }\href {https://arxiv.org/abs/2409.13504} {\bibfield
   {journal} {\bibinfo  {journal} {arXiv:2409.13504}\ } (\bibinfo {year}
  {2024})}\BibitemShut {NoStop}%
\bibitem [{\citenamefont {Smolyanyuk}\ \emph {et~al.}(2023)\citenamefont
  {Smolyanyuk}, \citenamefont {Mazin}, \citenamefont {Garcia-Gassull},\ and\
  \citenamefont {Valent{\'\i}}}]{Smolyanyuk2023arXiv}%
  \BibitemOpen
  \bibfield  {author} {\bibinfo {author} {\bibfnamefont {A.}~\bibnamefont
  {Smolyanyuk}}, \bibinfo {author} {\bibfnamefont {I.~I.}\ \bibnamefont
  {Mazin}}, \bibinfo {author} {\bibfnamefont {L.}~\bibnamefont
  {Garcia-Gassull}}, \ and\ \bibinfo {author} {\bibfnamefont {R.}~\bibnamefont
  {Valent{\'\i}}},\ }\bibfield  {title} {\enquote {\bibinfo {title} {{RuO$_2$:
  a puzzle to be solved}}}, }\href {https://arxiv.org/abs/2310.06909}
  {\bibfield  {journal} {\bibinfo  {journal} {arXiv:2310.06909}\ } (\bibinfo
  {year} {2023})}\BibitemShut {NoStop}%
\bibitem [{\citenamefont {Chang}\ \emph {et~al.}(2013)\citenamefont {Chang},
  \citenamefont {Zhang}, \citenamefont {Feng}, \citenamefont {Shen},
  \citenamefont {Zhang}, \citenamefont {Guo}, \citenamefont {Li}, \citenamefont
  {Ou}, \citenamefont {Wei}, \citenamefont {Wang} \emph
  {et~al.}}]{Chang2013Science}%
  \BibitemOpen
  \bibfield  {author} {\bibinfo {author} {\bibfnamefont {C.}~\bibnamefont
  {Chang}}, \bibinfo {author} {\bibfnamefont {J.}~\bibnamefont {Zhang}},
  \bibinfo {author} {\bibfnamefont {X.}~\bibnamefont {Feng}}, \bibinfo {author}
  {\bibfnamefont {J.}~\bibnamefont {Shen}}, \bibinfo {author} {\bibfnamefont
  {Z.}~\bibnamefont {Zhang}}, \bibinfo {author} {\bibfnamefont
  {M.}~\bibnamefont {Guo}},  \emph {et~al.},\ }\bibfield  {title} {\enquote
  {\bibinfo {title} {{Experimental observation of the quantum anomalous Hall
  effect in a magnetic topological insulator}}}, }\href {\doibase
  10.1126/science.1234414} {\bibfield  {journal} {\bibinfo  {journal}
  {Science}\ }\textbf {\bibinfo {volume} {340}},\ \bibinfo {pages} {167}
  (\bibinfo {year} {2013})}\BibitemShut {NoStop}%
\bibitem [{\citenamefont {Deng}\ \emph {et~al.}(2020)\citenamefont {Deng},
  \citenamefont {Yu}, \citenamefont {Shi}, \citenamefont {Guo}, \citenamefont
  {Xu}, \citenamefont {Wang}, \citenamefont {Chen},\ and\ \citenamefont
  {Zhang}}]{Deng20sci}%
  \BibitemOpen
  \bibfield  {author} {\bibinfo {author} {\bibfnamefont {Y.}~\bibnamefont
  {Deng}}, \bibinfo {author} {\bibfnamefont {Y.}~\bibnamefont {Yu}}, \bibinfo
  {author} {\bibfnamefont {M.~Z.}\ \bibnamefont {Shi}}, \bibinfo {author}
  {\bibfnamefont {Z.}~\bibnamefont {Guo}}, \bibinfo {author} {\bibfnamefont
  {Z.}~\bibnamefont {Xu}}, \bibinfo {author} {\bibfnamefont {J.}~\bibnamefont
  {Wang}}, \bibinfo {author} {\bibfnamefont {X.~H.}\ \bibnamefont {Chen}}, \
  and\ \bibinfo {author} {\bibfnamefont {Y.}~\bibnamefont {Zhang}},\ }\bibfield
   {title} {\enquote {\bibinfo {title} {Quantum anomalous {Hall} effect in
  intrinsic magnetic topological insulator {MnBi}$_2${Te}$_4$}}, }\href
  {https://science.sciencemag.org/content/367/6480/895} {\bibfield  {journal}
  {\bibinfo  {journal} {Science}\ }\textbf {\bibinfo {volume} {367}},\ \bibinfo
  {pages} {895} (\bibinfo {year} {2020})}\BibitemShut {NoStop}%
\bibitem [{\citenamefont {Mogi}\ \emph {et~al.}(2017)\citenamefont {Mogi},
  \citenamefont {Kawamura}, \citenamefont {Yoshimi}, \citenamefont {Tsukazaki},
  \citenamefont {Kozuka}, \citenamefont {Shirakawa}, \citenamefont {Takahashi},
  \citenamefont {Kawasaki},\ and\ \citenamefont {Tokura}}]{Mogi17nm}%
  \BibitemOpen
  \bibfield  {author} {\bibinfo {author} {\bibfnamefont {M.}~\bibnamefont
  {Mogi}}, \bibinfo {author} {\bibfnamefont {M.}~\bibnamefont {Kawamura}},
  \bibinfo {author} {\bibfnamefont {R.}~\bibnamefont {Yoshimi}}, \bibinfo
  {author} {\bibfnamefont {A.}~\bibnamefont {Tsukazaki}}, \bibinfo {author}
  {\bibfnamefont {Y.}~\bibnamefont {Kozuka}}, \bibinfo {author} {\bibfnamefont
  {N.}~\bibnamefont {Shirakawa}}, \bibinfo {author} {\bibfnamefont {K.~S.}\
  \bibnamefont {Takahashi}}, \bibinfo {author} {\bibfnamefont {M.}~\bibnamefont
  {Kawasaki}}, \ and\ \bibinfo {author} {\bibfnamefont {Y.}~\bibnamefont
  {Tokura}},\ }\bibfield  {title} {\enquote {\bibinfo {title} {A magnetic
  heterostructure of topological insulators as a candidate for an axion
  insulator}}, }\href {\doibase 10.1038/nmat4855} {\bibfield  {journal}
  {\bibinfo  {journal} {Nat. Mater.}\ }\textbf {\bibinfo {volume} {16}},\
  \bibinfo {pages} {516} (\bibinfo {year} {2017})}\BibitemShut {NoStop}%
\bibitem [{\citenamefont {Liu}\ \emph {et~al.}(2020)\citenamefont {Liu},
  \citenamefont {Wang}, \citenamefont {Li}, \citenamefont {Wu}, \citenamefont
  {Li}, \citenamefont {Li}, \citenamefont {He}, \citenamefont {Xu},
  \citenamefont {Zhang},\ and\ \citenamefont {Wang}}]{Liu20nm}%
  \BibitemOpen
  \bibfield  {author} {\bibinfo {author} {\bibfnamefont {C.}~\bibnamefont
  {Liu}}, \bibinfo {author} {\bibfnamefont {Y.}~\bibnamefont {Wang}}, \bibinfo
  {author} {\bibfnamefont {H.}~\bibnamefont {Li}}, \bibinfo {author}
  {\bibfnamefont {Y.}~\bibnamefont {Wu}}, \bibinfo {author} {\bibfnamefont
  {Y.}~\bibnamefont {Li}}, \bibinfo {author} {\bibfnamefont {J.}~\bibnamefont
  {Li}}, \bibinfo {author} {\bibfnamefont {K.}~\bibnamefont {He}}, \bibinfo
  {author} {\bibfnamefont {Y.}~\bibnamefont {Xu}}, \bibinfo {author}
  {\bibfnamefont {J.}~\bibnamefont {Zhang}}, \ and\ \bibinfo {author}
  {\bibfnamefont {Y.}~\bibnamefont {Wang}},\ }\bibfield  {title} {\enquote
  {\bibinfo {title} {Robust axion insulator and {Chern} insulator phases in a
  two-dimensional antiferromagnetic topological insulator}}, }\href
  {https://doi.org/10.1038/s41563-019-0573-3} {\bibfield  {journal} {\bibinfo
  {journal} {Nat. Mater.}\ }\textbf {\bibinfo {volume} {19}},\ \bibinfo {pages}
  {522} (\bibinfo {year} {2020})}\BibitemShut {NoStop}%
\bibitem [{\citenamefont {Mogi}\ \emph {et~al.}(2022)\citenamefont {Mogi},
  \citenamefont {Okamura}, \citenamefont {Kawamura}, \citenamefont {Yoshimi},
  \citenamefont {Yasuda}, \citenamefont {Tsukazaki}, \citenamefont {Takahashi},
  \citenamefont {Morimoto}, \citenamefont {Nagaosa}, \citenamefont {Kawasaki},
  \citenamefont {Takahashi},\ and\ \citenamefont
  {Tokura}}]{mogi2021experimental}%
  \BibitemOpen
  \bibfield  {author} {\bibinfo {author} {\bibfnamefont {M.}~\bibnamefont
  {Mogi}}, \bibinfo {author} {\bibfnamefont {Y.}~\bibnamefont {Okamura}},
  \bibinfo {author} {\bibfnamefont {M.}~\bibnamefont {Kawamura}}, \bibinfo
  {author} {\bibfnamefont {R.}~\bibnamefont {Yoshimi}}, \bibinfo {author}
  {\bibfnamefont {K.}~\bibnamefont {Yasuda}}, \bibinfo {author} {\bibfnamefont
  {A.}~\bibnamefont {Tsukazaki}},  \emph {et~al.},\ }\bibfield  {title}
  {\enquote {\bibinfo {title} {Experimental signature of the parity anomaly in
  a semi-magnetic topological insulator}}, }\href {\doibase
  10.1038/s41567-021-01490-y} {\bibfield  {journal} {\bibinfo  {journal} {Nat.
  Phys.}\ }\textbf {\bibinfo {volume} {18}},\ \bibinfo {pages} {390} (\bibinfo
  {year} {2022})}\BibitemShut {NoStop}%
\bibitem [{\citenamefont {Tokura}\ \emph {et~al.}(2019)\citenamefont {Tokura},
  \citenamefont {Yasuda},\ and\ \citenamefont {Tsukazaki}}]{Tokura19nrp}%
  \BibitemOpen
  \bibfield  {author} {\bibinfo {author} {\bibfnamefont {Y.}~\bibnamefont
  {Tokura}}, \bibinfo {author} {\bibfnamefont {K.}~\bibnamefont {Yasuda}}, \
  and\ \bibinfo {author} {\bibfnamefont {A.}~\bibnamefont {Tsukazaki}},\
  }\bibfield  {title} {\enquote {\bibinfo {title} {Magnetic topological
  insulators}}, }\href {https://doi.org/10.1038/s42254-018-0011-5} {\bibfield
  {journal} {\bibinfo  {journal} {Nat. Rev. Phys.}\ }\textbf {\bibinfo {volume}
  {1}},\ \bibinfo {pages} {126} (\bibinfo {year} {2019})}\BibitemShut {NoStop}%
\bibitem [{\citenamefont {Chang}(2020)}]{Chang2020NMRev}%
  \BibitemOpen
  \bibfield  {author} {\bibinfo {author} {\bibfnamefont {C.-Z.}\ \bibnamefont
  {Chang}},\ }\bibfield  {title} {\enquote {\bibinfo {title} {Marriage of
  topology and magnetism}}, }\href {\doibase 10.1038/s41563-020-0632-9}
  {\bibfield  {journal} {\bibinfo  {journal} {Nat. Mater.}\ }\textbf {\bibinfo
  {volume} {19}},\ \bibinfo {pages} {484–485} (\bibinfo {year}
  {2020})}\BibitemShut {NoStop}%
\bibitem [{\citenamefont {Wang}\ \emph {et~al.}(2023)\citenamefont {Wang},
  \citenamefont {Fu},\ and\ \citenamefont {Shen}}]{Wang2023}%
  \BibitemOpen
  \bibfield  {author} {\bibinfo {author} {\bibfnamefont {H.-W.}\ \bibnamefont
  {Wang}}, \bibinfo {author} {\bibfnamefont {B.}~\bibnamefont {Fu}}, \ and\
  \bibinfo {author} {\bibfnamefont {S.-Q.}\ \bibnamefont {Shen}},\ }\bibfield
  {title} {\enquote {\bibinfo {title} {Recent progress of transport theory in
  {Dirac} quantum materials}}, }\href {\doibase 10.7498/aps.72.20230672}
  {\bibfield  {journal} {\bibinfo  {journal} {Acta Phys. Sin-ch. Ed.}\ }\textbf
  {\bibinfo {volume} {72}},\ \bibinfo {pages} {177303} (\bibinfo {year}
  {2023})}\BibitemShut {NoStop}%
\bibitem [{\citenamefont {Fu}\ \emph {et~al.}(2007)\citenamefont {Fu},
  \citenamefont {Kane},\ and\ \citenamefont {Mele}}]{FL07PRL}%
  \BibitemOpen
  \bibfield  {author} {\bibinfo {author} {\bibfnamefont {L.}~\bibnamefont
  {Fu}}, \bibinfo {author} {\bibfnamefont {C.~L.}\ \bibnamefont {Kane}}, \ and\
  \bibinfo {author} {\bibfnamefont {E.~J.}\ \bibnamefont {Mele}},\ }\bibfield
  {title} {\enquote {\bibinfo {title} {Topological insulators in three
  dimensions}}, }\href {\doibase 10.1103/PhysRevLett.98.106803} {\bibfield
  {journal} {\bibinfo  {journal} {Phys. Rev. Lett.}\ }\textbf {\bibinfo
  {volume} {98}},\ \bibinfo {pages} {106803} (\bibinfo {year}
  {2007})}\BibitemShut {NoStop}%
\bibitem [{\citenamefont {Hasan}\ and\ \citenamefont
  {Kane}(2010)}]{Hasan2010RMP}%
  \BibitemOpen
  \bibfield  {author} {\bibinfo {author} {\bibfnamefont {M.~Z.}\ \bibnamefont
  {Hasan}}\ and\ \bibinfo {author} {\bibfnamefont {C.~L.}\ \bibnamefont
  {Kane}},\ }\bibfield  {title} {\enquote {\bibinfo {title} {{Colloquium:
  Topological insulators}}}, }\href {\doibase 10.1103/revmodphys.82.3045}
  {\bibfield  {journal} {\bibinfo  {journal} {Rev. Mod. Phys.}\ }\textbf
  {\bibinfo {volume} {82}},\ \bibinfo {pages} {3045} (\bibinfo {year}
  {2010})}\BibitemShut {NoStop}%
\bibitem [{\citenamefont {Liu}\ \emph {et~al.}(2010)\citenamefont {Liu},
  \citenamefont {Qi}, \citenamefont {Zhang}, \citenamefont {Dai}, \citenamefont
  {Fang},\ and\ \citenamefont {Zhang}}]{Liu10prb}%
  \BibitemOpen
  \bibfield  {author} {\bibinfo {author} {\bibfnamefont {C.-X.}\ \bibnamefont
  {Liu}}, \bibinfo {author} {\bibfnamefont {X.-L.}\ \bibnamefont {Qi}},
  \bibinfo {author} {\bibfnamefont {H.~J.}\ \bibnamefont {Zhang}}, \bibinfo
  {author} {\bibfnamefont {X.}~\bibnamefont {Dai}}, \bibinfo {author}
  {\bibfnamefont {Z.}~\bibnamefont {Fang}}, \ and\ \bibinfo {author}
  {\bibfnamefont {S.-C.}\ \bibnamefont {Zhang}},\ }\bibfield  {title} {\enquote
  {\bibinfo {title} {Model {Hamiltonian} for topological insulators}}, }\href
  {\doibase 10.1103/PhysRevB.82.045122} {\bibfield  {journal} {\bibinfo
  {journal} {Phys. Rev. B}\ }\textbf {\bibinfo {volume} {82}},\ \bibinfo
  {pages} {045122} (\bibinfo {year} {2010})}\BibitemShut {NoStop}%
\bibitem [{\citenamefont {Zhang}\ \emph {et~al.}(2009)\citenamefont {Zhang},
  \citenamefont {Liu}, \citenamefont {Qi}, \citenamefont {Dai}, \citenamefont
  {Fang},\ and\ \citenamefont {Zhang}}]{Zhang09np}%
  \BibitemOpen
  \bibfield  {author} {\bibinfo {author} {\bibfnamefont {H.}~\bibnamefont
  {Zhang}}, \bibinfo {author} {\bibfnamefont {C.-X.}\ \bibnamefont {Liu}},
  \bibinfo {author} {\bibfnamefont {X.-L.}\ \bibnamefont {Qi}}, \bibinfo
  {author} {\bibfnamefont {X.}~\bibnamefont {Dai}}, \bibinfo {author}
  {\bibfnamefont {Z.}~\bibnamefont {Fang}}, \ and\ \bibinfo {author}
  {\bibfnamefont {S.-C.}\ \bibnamefont {Zhang}},\ }\bibfield  {title} {\enquote
  {\bibinfo {title} {Topological insulators in {Bi}$_2${Se}$_3$,
  {Bi}$_2${Te}$_3$ and {Sb}$_2${Te}$_3$ with a single {Dirac} cone on the
  surface}}, }\href {\doibase 10.1038/NPHYS1270} {\bibfield  {journal}
  {\bibinfo  {journal} {Nature Phys.}\ }\textbf {\bibinfo {volume} {5}},\
  \bibinfo {pages} {438} (\bibinfo {year} {2009})}\BibitemShut {NoStop}%
\bibitem [{\citenamefont {Liu}\ \emph {et~al.}(2013)\citenamefont {Liu},
  \citenamefont {Hsu},\ and\ \citenamefont {Liu}}]{Liu2013}%
  \BibitemOpen
  \bibfield  {author} {\bibinfo {author} {\bibfnamefont {X.}~\bibnamefont
  {Liu}}, \bibinfo {author} {\bibfnamefont {H.-C.}\ \bibnamefont {Hsu}}, \ and\
  \bibinfo {author} {\bibfnamefont {C.-X.}\ \bibnamefont {Liu}},\ }\bibfield
  {title} {\enquote {\bibinfo {title} {In-plane magnetization-induced quantum
  anomalous {Hall} effect}}, }\href {\doibase 10.1103/PhysRevLett.111.086802}
  {\bibfield  {journal} {\bibinfo  {journal} {Phys. Rev. Lett.}\ }\textbf
  {\bibinfo {volume} {111}},\ \bibinfo {pages} {086802} (\bibinfo {year}
  {2013})}\BibitemShut {NoStop}%
\bibitem [{\citenamefont {Yu}\ \emph {et~al.}(2010)\citenamefont {Yu},
  \citenamefont {Zhang}, \citenamefont {Zhang}, \citenamefont {Zhang},
  \citenamefont {Dai},\ and\ \citenamefont {Fang}}]{Yu2010Science}%
  \BibitemOpen
  \bibfield  {author} {\bibinfo {author} {\bibfnamefont {R.}~\bibnamefont
  {Yu}}, \bibinfo {author} {\bibfnamefont {W.}~\bibnamefont {Zhang}}, \bibinfo
  {author} {\bibfnamefont {H.-J.}\ \bibnamefont {Zhang}}, \bibinfo {author}
  {\bibfnamefont {S.-C.}\ \bibnamefont {Zhang}}, \bibinfo {author}
  {\bibfnamefont {X.}~\bibnamefont {Dai}}, \ and\ \bibinfo {author}
  {\bibfnamefont {Z.}~\bibnamefont {Fang}},\ }\bibfield  {title} {\enquote
  {\bibinfo {title} {Quantized anomalous {Hall} effect in magnetic topological
  insulators}}, }\href {\doibase 10.1126/science.1187485} {\bibfield  {journal}
  {\bibinfo  {journal} {Science}\ }\textbf {\bibinfo {volume} {329}},\ \bibinfo
  {pages} {61} (\bibinfo {year} {2010})}\BibitemShut {NoStop}%
\bibitem [{\citenamefont {Kandala}\ \emph {et~al.}(2015)\citenamefont
  {Kandala}, \citenamefont {Richardella}, \citenamefont {Kempinger},
  \citenamefont {Liu},\ and\ \citenamefont {Samarth}}]{Kandala2015NC}%
  \BibitemOpen
  \bibfield  {author} {\bibinfo {author} {\bibfnamefont {A.}~\bibnamefont
  {Kandala}}, \bibinfo {author} {\bibfnamefont {A.}~\bibnamefont
  {Richardella}}, \bibinfo {author} {\bibfnamefont {S.}~\bibnamefont
  {Kempinger}}, \bibinfo {author} {\bibfnamefont {C.-X.}\ \bibnamefont {Liu}},
  \ and\ \bibinfo {author} {\bibfnamefont {N.}~\bibnamefont {Samarth}},\
  }\bibfield  {title} {\enquote {\bibinfo {title} {Giant anisotropic
  magnetoresistance in a quantum anomalous {Hall} insulator}}, }\href {\doibase
  10.1038/ncomms8434} {\bibfield  {journal} {\bibinfo  {journal} {Nat.
  Commun.}\ }\textbf {\bibinfo {volume} {6}},\ \bibinfo {pages} {7434}
  (\bibinfo {year} {2015})}\BibitemShut {NoStop}%
\bibitem [{\citenamefont {Tschirner}\ \emph {et~al.}(2023)\citenamefont
  {Tschirner}, \citenamefont {Keßler}, \citenamefont {Gonzalez~Betancourt},
  \citenamefont {Kotte}, \citenamefont {Kriegner}, \citenamefont {Büchner},
  \citenamefont {Dufouleur}, \citenamefont {Kamp}, \citenamefont {Jovic},
  \citenamefont {Smejkal}, \citenamefont {Sinova}, \citenamefont {Claessen},
  \citenamefont {Jungwirth}, \citenamefont {Moser}, \citenamefont {Reichlova},\
  and\ \citenamefont {Veyrat}}]{Tschirner2023APL}%
  \BibitemOpen
  \bibfield  {author} {\bibinfo {author} {\bibfnamefont {T.}~\bibnamefont
  {Tschirner}}, \bibinfo {author} {\bibfnamefont {P.}~\bibnamefont {Keßler}},
  \bibinfo {author} {\bibfnamefont {R.~D.}\ \bibnamefont
  {Gonzalez~Betancourt}}, \bibinfo {author} {\bibfnamefont {T.}~\bibnamefont
  {Kotte}}, \bibinfo {author} {\bibfnamefont {D.}~\bibnamefont {Kriegner}},
  \bibinfo {author} {\bibfnamefont {B.}~\bibnamefont {Büchner}},  \emph
  {et~al.},\ }\bibfield  {title} {\enquote {\bibinfo {title} {Saturation of the
  anomalous hall effect at high magnetic fields in altermagnetic ruo2}}, }\href
  {\doibase 10.1063/5.0160335} {\bibfield  {journal} {\bibinfo  {journal} {APL
  Mater.}\ }\textbf {\bibinfo {volume} {11}},\ \bibinfo {pages} {101103}
  (\bibinfo {year} {2023})}\BibitemShut {NoStop}%
\bibitem [{\citenamefont {Gonzalez~Betancourt}\ \emph
  {et~al.}(2023)\citenamefont {Gonzalez~Betancourt}, \citenamefont
  {Zub\'a\ifmmode~\check{c}\else \v{c}\fi{}}, \citenamefont
  {Gonzalez-Hernandez}, \citenamefont {Geishendorf}, \citenamefont {\ifmmode
  \check{S}\else \v{S}\fi{}ob\'a\ifmmode~\check{n}\else \v{n}\fi{}},
  \citenamefont {Springholz}, \citenamefont {Olejn\'{\i}k}, \citenamefont
  {\ifmmode~\check{S}\else \v{S}\fi{}mejkal}, \citenamefont {Sinova},
  \citenamefont {Jungwirth}, \citenamefont {Goennenwein}, \citenamefont
  {Thomas}, \citenamefont {Reichlov\'a}, \citenamefont {\ifmmode~\check{Z}\else
  \v{Z}\fi{}elezn\'y},\ and\ \citenamefont {Kriegner}}]{Gonzalez2023PRL}%
  \BibitemOpen
  \bibfield  {author} {\bibinfo {author} {\bibfnamefont {R.~D.}\ \bibnamefont
  {Gonzalez~Betancourt}}, \bibinfo {author} {\bibfnamefont {J.}~\bibnamefont
  {Zub\'a\ifmmode~\check{c}\else \v{c}\fi{}}}, \bibinfo {author} {\bibfnamefont
  {R.}~\bibnamefont {Gonzalez-Hernandez}}, \bibinfo {author} {\bibfnamefont
  {K.}~\bibnamefont {Geishendorf}}, \bibinfo {author} {\bibfnamefont
  {Z.}~\bibnamefont {\ifmmode \check{S}\else
  \v{S}\fi{}ob\'a\ifmmode~\check{n}\else \v{n}\fi{}}}, \bibinfo {author}
  {\bibfnamefont {G.}~\bibnamefont {Springholz}},  \emph {et~al.},\ }\bibfield
  {title} {\enquote {\bibinfo {title} {Spontaneous anomalous hall effect
  arising from an unconventional compensated magnetic phase in a
  semiconductor}}, }\href {\doibase 10.1103/PhysRevLett.130.036702} {\bibfield
  {journal} {\bibinfo  {journal} {Phys. Rev. Lett.}\ }\textbf {\bibinfo
  {volume} {130}},\ \bibinfo {pages} {036702} (\bibinfo {year}
  {2023})}\BibitemShut {NoStop}%
\bibitem [{Sup()}]{Supp}%
  \BibitemOpen
  \href@noop {} {\bibinfo  {journal} {See Supplemental Material for more
  details}\ }\BibitemShut {NoStop}%
\bibitem [{\citenamefont {Shen}(2017)}]{Shen2017TI}%
  \BibitemOpen
\bibfield  {journal} {  }\bibfield  {author} {\bibinfo {author} {\bibfnamefont
  {S.-Q.}\ \bibnamefont {Shen}},\ }\href {\doibase 10.1007/978-981-10-4606-3}
  {\emph {\bibinfo {title} {Topological Insulators--Dirac Equation in Condensed
  Matter}}}\ (\bibinfo  {publisher} {Springer Singapore},\ \bibinfo {year}
  {2017})\BibitemShut {NoStop}%
\bibitem [{\citenamefont {Qi}\ and\ \citenamefont {Zhang}(2011)}]{Qi2011RMP}%
  \BibitemOpen
  \bibfield  {author} {\bibinfo {author} {\bibfnamefont {X.-L.}\ \bibnamefont
  {Qi}}\ and\ \bibinfo {author} {\bibfnamefont {S.-C.}\ \bibnamefont {Zhang}},\
  }\bibfield  {title} {\enquote {\bibinfo {title} {{Topological insulators and
  superconductors}}}, }\href {\doibase 10.1103/revmodphys.83.1057} {\bibfield
  {journal} {\bibinfo  {journal} {Rev. Mod. Phys.}\ }\textbf {\bibinfo {volume}
  {83}},\ \bibinfo {pages} {1057} (\bibinfo {year} {2011})}\BibitemShut
  {NoStop}%
\bibitem [{\citenamefont {Fu}\ \emph {et~al.}(2022)\citenamefont {Fu},
  \citenamefont {Zou}, \citenamefont {Hu}, \citenamefont {Wang},\ and\
  \citenamefont {Shen}}]{Fu2022NPJQM}%
  \BibitemOpen
  \bibfield  {author} {\bibinfo {author} {\bibfnamefont {B.}~\bibnamefont
  {Fu}}, \bibinfo {author} {\bibfnamefont {J.-Y.}\ \bibnamefont {Zou}},
  \bibinfo {author} {\bibfnamefont {Z.-A.}\ \bibnamefont {Hu}}, \bibinfo
  {author} {\bibfnamefont {H.-W.}\ \bibnamefont {Wang}}, \ and\ \bibinfo
  {author} {\bibfnamefont {S.-Q.}\ \bibnamefont {Shen}},\ }\bibfield  {title}
  {\enquote {\bibinfo {title} {Quantum anomalous semimetals}}, }\href {\doibase
  10.1038/s41535-022-00503-0} {\bibfield  {journal} {\bibinfo  {journal} {npj
  Quantum Materials}\ }\textbf {\bibinfo {volume} {7}},\ \bibinfo {pages} {94}
  (\bibinfo {year} {2022})}\BibitemShut {NoStop}%
\bibitem [{\citenamefont {Landauer}(1970)}]{Landauer1970Philosophical}%
  \BibitemOpen
  \bibfield  {author} {\bibinfo {author} {\bibfnamefont {R.}~\bibnamefont
  {Landauer}},\ }\bibfield  {title} {\enquote {\bibinfo {title} {{Electrical
  resistance of disordered one-dimensional lattices}}}, }\href {\doibase
  10.1080/14786437008238472} {\bibfield  {journal} {\bibinfo  {journal}
  {Philos. Mag.}\ }\textbf {\bibinfo {volume} {21}},\ \bibinfo {pages} {863}
  (\bibinfo {year} {1970})}\BibitemShut {NoStop}%
\bibitem [{\citenamefont {B\"uttiker}(1988)}]{Buttiker1988PRB}%
  \BibitemOpen
  \bibfield  {author} {\bibinfo {author} {\bibfnamefont {M.}~\bibnamefont
  {B\"uttiker}},\ }\bibfield  {title} {\enquote {\bibinfo {title} {{Absence of
  backscattering in the quantum Hall effect in multiprobe conductors}}}, }\href
  {\doibase 10.1103/physrevb.38.9375} {\bibfield  {journal} {\bibinfo
  {journal} {Phys. Rev. B}\ }\textbf {\bibinfo {volume} {38}},\ \bibinfo
  {pages} {9375} (\bibinfo {year} {1988})}\BibitemShut {NoStop}%
\bibitem [{\citenamefont {Fisher}\ and\ \citenamefont
  {Lee}(1981)}]{Fisher1981PRB}%
  \BibitemOpen
  \bibfield  {author} {\bibinfo {author} {\bibfnamefont {D.~S.}\ \bibnamefont
  {Fisher}}\ and\ \bibinfo {author} {\bibfnamefont {P.~A.}\ \bibnamefont
  {Lee}},\ }\bibfield  {title} {\enquote {\bibinfo {title} {{Relation between
  conductivity and transmission matrix}}}, }\href {\doibase
  10.1103/physrevb.23.6851} {\bibfield  {journal} {\bibinfo  {journal} {Phys.
  Rev. B}\ }\textbf {\bibinfo {volume} {23}},\ \bibinfo {pages} {6851}
  (\bibinfo {year} {1981})}\BibitemShut {NoStop}%
\bibitem [{\citenamefont {MacKinnon}(1985)}]{Mackinnon1985Zeitschrift}%
  \BibitemOpen
  \bibfield  {author} {\bibinfo {author} {\bibfnamefont {A.}~\bibnamefont
  {MacKinnon}},\ }\bibfield  {title} {\enquote {\bibinfo {title} {{The
  calculation of transport properties and density of states of disordered
  solids}}}, }\href {\doibase 10.1007/bf01328846} {\bibfield  {journal}
  {\bibinfo  {journal} {Z. Phys. B}\ }\textbf {\bibinfo {volume} {59}},\
  \bibinfo {pages} {385} (\bibinfo {year} {1985})}\BibitemShut {NoStop}%
\bibitem [{\citenamefont {Metalidis}\ and\ \citenamefont
  {Bruno}(2005)}]{Metalidis2005PRB}%
  \BibitemOpen
  \bibfield  {author} {\bibinfo {author} {\bibfnamefont {G.}~\bibnamefont
  {Metalidis}}\ and\ \bibinfo {author} {\bibfnamefont {P.}~\bibnamefont
  {Bruno}},\ }\bibfield  {title} {\enquote {\bibinfo {title} {{Green's function
  technique for studying electron flow in two-dimensional mesoscopic
  samples}}}, }\href {\doibase 10.1103/physrevb.72.235304} {\bibfield
  {journal} {\bibinfo  {journal} {Phys. Rev. B}\ }\textbf {\bibinfo {volume}
  {72}},\ \bibinfo {pages} {235304} (\bibinfo {year} {2005})}\BibitemShut
  {NoStop}%
\bibitem [{\citenamefont {Fu}(2009)}]{FuL09PRL}%
  \BibitemOpen
  \bibfield  {author} {\bibinfo {author} {\bibfnamefont {L.}~\bibnamefont
  {Fu}},\ }\bibfield  {title} {\enquote {\bibinfo {title} {{Hexagonal Warping
  Effects in the Surface States of the Topological Insulator
  ${\mathrm{Bi}}_{2}{\mathrm{Te}}_{3}$}}}, }\href {\doibase
  10.1103/PhysRevLett.103.266801} {\bibfield  {journal} {\bibinfo  {journal}
  {Phys. Rev. Lett.}\ }\textbf {\bibinfo {volume} {103}},\ \bibinfo {pages}
  {266801} (\bibinfo {year} {2009})}\BibitemShut {NoStop}%
\bibitem [{\citenamefont {Leivisk\"a}\ \emph {et~al.}(2024)\citenamefont
  {Leivisk\"a}, \citenamefont {Rial}, \citenamefont {Bad'ura}, \citenamefont
  {Seeger}, \citenamefont {Kounta}, \citenamefont {Beckert}, \citenamefont
  {Kriegner}, \citenamefont {Joumard}, \citenamefont {Schmoranzerov\'a},
  \citenamefont {Sinova}, \citenamefont {Gomonay}, \citenamefont {Thomas},
  \citenamefont {Goennenwein}, \citenamefont {Reichlov\'a}, \citenamefont
  {\ifmmode~\check{S}\else \v{S}\fi{}mejkal}, \citenamefont {Michez},
  \citenamefont {Jungwirth},\ and\ \citenamefont {Baltz}}]{Leiviska2024PRB}%
  \BibitemOpen
  \bibfield  {author} {\bibinfo {author} {\bibfnamefont {M.}~\bibnamefont
  {Leivisk\"a}}, \bibinfo {author} {\bibfnamefont {J.}~\bibnamefont {Rial}},
  \bibinfo {author} {\bibfnamefont {A.}~\bibnamefont {Bad'ura}}, \bibinfo
  {author} {\bibfnamefont {R.~L.}\ \bibnamefont {Seeger}}, \bibinfo {author}
  {\bibfnamefont {I.}~\bibnamefont {Kounta}}, \bibinfo {author} {\bibfnamefont
  {S.}~\bibnamefont {Beckert}},  \emph {et~al.},\ }\bibfield  {title} {\enquote
  {\bibinfo {title} {Anisotropy of the anomalous hall effect in thin films of
  the altermagnet candidate ${\mathrm{mn}}_{5}{\mathrm{si}}_{3}$}}, }\href
  {\doibase 10.1103/PhysRevB.109.224430} {\bibfield  {journal} {\bibinfo
  {journal} {Phys. Rev. B}\ }\textbf {\bibinfo {volume} {109}},\ \bibinfo
  {pages} {224430} (\bibinfo {year} {2024})}\BibitemShut {NoStop}%
\end{thebibliography}%

\end{document}